\documentclass[showpacs,amsmath,twocolumn,showkeys,pra]{revtex4-1}
\usepackage[T1]{fontenc}
\usepackage{inputenc}
\usepackage[margin=1in]{geometry}
\usepackage{color,times}
\usepackage{array}
\usepackage{amsmath}
\usepackage{stackrel}
\usepackage{graphicx}
\usepackage{appendix}
\usepackage{mathrsfs}
\usepackage{amsfonts}
\usepackage{appendix}
\usepackage{hyperref}
\usepackage{color}
\definecolor{med-blue}{RGB}{25,25,112} 
\hypersetup{colorlinks, linkcolor={red},citecolor={blue}, urlcolor={blue}}
\allowdisplaybreaks
\newcommand{\ket}[1]{\vert{#1}\rangle}
\newcommand{\bra}[1]{\langle{#1}\vert}

\newcommand{\expv}[1]{\langle{#1}\rangle}

\newcommand{\tr}{\mathrm{Tr}}

\providecommand{\tabularnewline}{\\}

\begin{document}

\title
{Complete characterization of the directly implementable quantum gates used in the IBM quantum processors}

\author{Abhishek Shukla, Mitali Sisodia, and Anirban
 Pathak}
\affiliation{Jaypee Institute of Information Technology, A 10, Sector 62, Noida, UP 201307, India}
 \affiliation{University of Science and Technology of China, Hefei, 230026, P. R. China}

\begin{abstract}
Quantum process tomography of  each directly implementable quantum gate used in the IBM quantum processors is performed  to  compute gate error in order to check viability of complex quantum operations in the superconductivity-based quantum computers introduced by  IBM and to compare the quality of these gates with the corresponding gates implemented using other technologies. Quantum process tomography (QPT) of C-NOT gates have been performed for three configurations available in IBM QX4 processor. For all the other allowed gates QPT have been performed  for every allowed position (i.e., by placing the gates in different qubit lines) for IBM QX4 architecture, and thus, gate fidelities are obtained for both single-qubit and 2-qubit gates. Gate fidelities are observed to be lower than the corresponding values obtained  in the other technologies, like NMR. Further, gate fidelities for all the single-qubit  gates are obtained for IBM QX2 architecture by placing the gates in the third qubit line ($q[2]$).  It's observed that the IBM QX4 architecture yields better gate fidelity compared to IBM QX2 in all cases except the case of $\operatorname{Y}$ gate as far as the gate fidelity corresponding to the third qubit line is concerned. In general, the analysis performed here leads to a conclusion that a considerable technological improvement would be inevitable to achieve the desired scalability required for the realization of complex quantum operations.

\end{abstract}

\keywords{state tomography, process tomography, IBM quantum experience, quantum process tomography, gate fidelity}
\pacs{03.67.Lx, 03.67.Ac, 03.65.Wj, 03.65.Ta}
\maketitle

\section{Introduction} \label{intro}
Experimental implementation of any quantum information processing task is all about initializing the system qubits to an aptly chosen quantum state followed by manipulating the initial state through a set of unitary and/or non-unitary quantum operations to obtain the desired output state and ultimately measuring that to obtain the expectation value of the desired observable. 
Implementation of a multi-qubit unitary quantum operation of arbitrary dimension requires application of local and non-local operations (quantum gates). Local operations are usually realized by means of external control fields, whereas the  non-local operations are implemented by exploiting internal Hamiltonian. Perfect realization of these quantum operations is possible only in the ideal scenario. In practice, the accuracy with which a quantum operation can be realized is limited by imperfect designing and implementation errors, and environmental effects. To circumvent or to  suppress these factors which reduces accuracy, we need to design robust control fields. In order to design control fields robust against environmental effects, design errors and implementation errors, it becomes imperative to characterize the control fields. The procedure of characterizing quantum operation and hence control fields is known as quantum process tomography (QPT) \cite{chuang97,zollar97}.  It  (i.e., QPT) inherently involves QST \cite{ChuangPRL98}. This is so because a quantum operation (except the selective measurements) linearly  maps input density matrix to output density matrix. The procedure of QPT involves initializing system-qubits in an aptly chosen density matrix basis followed by application of a process on these basis states and ultimately characterizing output quantum states through QST.

The idea of QPT was introduced in 1997, but the first experimental demonstration of QPT was done in 2001 using a nuclear magnetic resonance- (NMR) based architecture \cite{ChuangPRA2001,QPTofQFT}. Since then QPT has been performed using various quantum computing  architectures, including  architectures based on linear optics \cite{EAPT1Exp,altepeter,obrian,bellstatefilter}, ion traps \cite{qptITprl2006,QPT_IonTrapNature2010}, superconductivity
\cite{QPT_SQUID,SQUID2009,MartiniSQUID2010,QPT2spSQUID,QPT_SQUIDChow2011,SQUID_Dewes2012}, and
nitrogen vacancy (NV) center  \cite{suterprotectedgate,howard2006quantum}. Until now, a large number of modified and new methods for QPT have also been devised. Most common methods are like ancilla-assisted quantum process tomography (AAQPT) which reduces the number of experiments required to be done for the complete characterization of a quantum operation by exploiting ancillary qubits \cite{altepeter}, single-scan quantum process tomography which allows characterization of a process in one scan and thus allows to tomograph a dynamical processes \cite{sspt}, adaptive process tomography which utilizes the advantages of the adaptive strategies applied to state reconstruction, and allows to tomograph a process with ultimate bound of precision for state tomography \cite{pogorelov2017experimental}. These schemes for QPT are demanding in the sense that either they require to perform a large number of experiments or they require a large amount of resources \cite{chuang97,altepeter,sspt}. This is why a set  of simplified QPT schemes have been developed. Such a scheme exploits prior knowledge of type of interaction involved in generating the dynamics and hence remove scaling problem involved in a standard method for QPT \cite{simplifiedQPT1,simplifiedQPT2} and that in the recently devised bootstrap tomography \cite{dobrovitski2010bootstrap,gaikwad}. Another method which is based on error detection and does not involve QST is direct characterization of quantum dynamics \cite{mohseni2006direct}.

In the last few years, QPT has played an extremely important role in the experimental quantum computing and communication, as it has led to the procedures for  designing more precise and robust control fields. This is what motivated us to perform QPT for the quantum operations that are implemented in the different architectures of 5 qubit IBM quantum computers (i.e., in IBM QX2 and IBM QX4). Here, it may be apt to note that the IBM Corporation has recently introduced a quantum computing platform \cite{IBMQXE} with two five-qubit quantum processors \cite{IBMQX} accessible through the cloud to the registered users. Quantum information processing (QIP) on these processors involves initialization, manipulation, and measurement of Transmon qubits by utilizing superconductivity-based QIP architecture. A QIP task in an IBM quantum processor may involve a set of unitary and non-unitary operations (measurements). Unitary operations in the IBM quantum computers is governed by a quantum circuit comprised of single-qubit and multi-qubit gates selected from the Clifford+T gate library. In order to examine the goodness of these gates, in what follows, we have completely characterized the quantum gates using standard method of QPT. 
This investigation is further motivated by the fact that the IBM quantum computers have already been utilized to implement various quantum computing (e.g., \cite{devitt2016performing,wang201816,kandala123}) and communication \cite{dall2018device,sisodia2017design} tasks and in many of these works quantum state tomography (QST) has been performed and quantum state fidelity  has been reported to be low or moderate \cite{abc,satyajit2017discrimination}. This is in sharp contrast to the fact that the fidelities of the gates given in the IBM quantum processors' backend information \cite{backendQx2} is very high. Interestingly, no serious effort has yet been made to perform QPT of these gates and to investigate the origin of low or moderate quantum state fidelity as a consequence of the imperfection of the implemented gates (i.e., low gate fidelity). Present work aims to look into this particular perspective, and thus to establish that the present technology would require considerable improvement in gate fidelity to achieve the desired scalability.
 
The rest of the paper is organized in the following way. In Sec. \ref{theory}, we briefly describe the theory of QPT. In Sec. \ref{exp}, we present a scheme for performing QPT on IBM QX4 and IBM QX2 quantum processors. Our results of QPT performed on IBM quantum processors are presented and analyzed in Sec. \ref{results}. The paper is finally concluded in Sec. \ref{conclusions}

\section{Theory of QPT} \label{theory}

A general quantum operation  $\epsilon(\rho)$, linearly maps a density matrix $\rho$ into another density matrix $\rho\prime $ such that $\rho\prime=\epsilon(\rho)$. Here, $\rho$ is the input density matrix and $\rho\prime$ is the output density matrix. In the following, for the sake of completeness, we briefly describe the theoretical framework of QPT \cite{chuang97,zollar97,chuangbook}. In operator sum representation, output density matrix $\rho\prime$ can be written as 

\begin{equation}
\epsilon(\rho) = \sum_k E_{k} \rho E_{k}^{\dagger}, 
\end{equation}
where, $\{E_{k}\}$ are Kraus operators such that $\sum_{k} E^{\dagger}_{k}E_{k}= \operatorname{\mathbb{I}}$ \cite{chuangbook}. Determining quantum operation $\epsilon$ is equivalent to determining a set of operational elements $\{E_{k}\}$. In practice, determining these operators correspond to determining a set of complex numbers for each operational element $E_{k}$. Thus, expanding $E_{k}$ in terms of fixed set of operators $\{\tilde{E_m}\}$ such as $E_k = \sum_{m} e_{km} \tilde{E_m}$, where $\tilde{E_m}$ forms an operator basis. 
\begin{equation}
\epsilon(\rho) = \sum_{(m,n)} \chi_{mn} \tilde{E_m} \rho {\tilde{E_n}}^\dagger.
\end{equation}
Here, $\chi_{mn}= e_{km}{e_{km}}^{*}$ are elements of a positive Hermitian matrix known as $\chi$ matrix. Matrix $\chi$ completely characterizes the corresponding quantum process expanded in operator basis $\{\tilde{E_m\}}$. This is why $\chi$ matrix is also known as process matrix.  Let's choose \{$\rho_{j}\}$ as the basis in which a density matrix of arbitrary dimension $d=2^n$ can be expanded, where $1 \le j \le d^2$. Then,
\begin{equation}
\epsilon(\rho) = \sum_{(m,n)} \sum_{j} \chi_{mn} \tilde{E_m} \rho_j {\tilde{E_n}}^\dagger. \label{fixedset}
\end{equation}
and $\epsilon (\rho_j)$ can be written as 
\begin{equation}
\epsilon(\rho_j) = \sum_{k} \lambda_{jk}\rho_{k}. \label{lambda}
\end{equation}
Here, $\epsilon (\rho_j)$ is the experimental outcome obtained from the state tomography performed for the input state $\rho_{j}$ and allows us to compute $\lambda_{jk}$, also comparison of Eq. \ref{fixedset} and Eq. \ref{lambda} leads to the relation 
\begin{equation}
\lambda_{jk} = \sum_{mn} \chi_{mn} \beta^{jk}_{mn}.  \label{chi}
\end{equation}
For given input density matrix $\rho_{j}$, output density matrix $\rho_{k}$, and basis operator $\tilde{E_{m}}$ $\beta^{jk}_{mn}$ can be calculated using 
\begin{equation}
 \sum_{k} \beta^{jk}_{mn} \rho_{k} = \tilde{E_m} \rho_j {\tilde{E_n}}^\dagger. 
 \end{equation} \label{beta}
  While, $\lambda_{jk}$ can be obtained from tomography results using Eq. \ref{lambda}. As inversion of $\beta^{jk}_{mn}$  is guaranteed (see Ref. \cite{chuangbook}), one can rewrite Eq. \ref{chi} as follows.
\begin{equation}
\chi =  \beta^{-1} \lambda.  \label{chi1}
\end{equation}
Here, $\chi$ and $\lambda$ are column vectors of dimension $d^4 \times 1$ containing elements $\chi_{mn}$ and $\lambda_{jk}$ respectively while $\beta^{-1}$ is a matrix of dimension $d^4 \times d^4$. Matrix $\beta$ is then constructed by arranging elements $\beta^{jk}_{mn}$ for $jk$ row-wise and $mn$ column-wise. Keeping this theoretical framework in mind, in the following section, we will describe an explicit protocol for performing experimental QPT on the IBM quantum processors.

\section{Experimental scheme for performing QPT on the IBM quantum processors} \label{exp}
IBM quantum processors are based on superconducting Transmon qubits. Our experiments have been executed on IBM QX2 and IBM QX4 processors. Topology of superconducting qubits and allowed C-NOT operations in each of these architectures are given in \cite{IBMQX}. Description of the architectures topology, type of qubits, means to manipulate them (control fields, transmission lines, and gate library), and measurement schemes have been provided in \cite{backendQx2,backendQx4}.  A useful summary of the architecture description is also available in \cite{abc,sisodia2017design}. Here we restrict ourselves from restating those information. However, for the sake of completeness, we provide important experimental parameters including Hamiltonian parameters and decoherence times for IBM QX2 and IBM QX4 processors in Table \ref{exppar1} and in Table \ref{exppar2}, respectively. This is in accordance with the information provided by IBM in \cite{backendQx2,backendQx4}. 

\begin{widetext}

\begin{table}[h]
	
\begin{center}
	\begin{tabular}{|>{\centering}p{2cm}|>{\centering}p{1.5cm}|>{\centering}p{1.5cm}|>{\centering}p{1.5cm}|>{\centering}p{1.5cm}|>{\centering}p{1.5cm}|}
			\hline 
		    Qubit number q[i] & q[0]&q[1]&q[2]&q[3]&q[4] \tabularnewline
			\hline
			$\omega^{R}/2\pi(\operatorname{GHz})$&6.530350&6.481848&6.436229&6.579431&6.530225 \tabularnewline
			\hline
       	$\omega/2\pi(\operatorname{GHz})$&5.2723&5.2145&5.0289&5.2971&5.0561 \tabularnewline
			\hline 
		$\delta/2\pi(\operatorname{MHz})$&-330.3&-331.9&-331.2&-329.4&-335.5 \tabularnewline
			\hline 
		$\textbf{$\chi$}/2\pi(\operatorname{kHz})$&476&395&428&412&339 \tabularnewline 
			\hline 
		$\kappa/2\pi(\operatorname{kHz})$&523&489&415&515&480 \tabularnewline
		\hline 	
		$\operatorname T_{1}(\mu \operatorname s)$ &53.04 &63.94 &52.08 &51.78 &55.80  \tabularnewline
		\hline
		$\operatorname T_{2}(\mu \operatorname s)$ &48.50 &35.07 &89.73 &60.93 &84.18  \tabularnewline	
		\hline
		\end{tabular} 
		\end{center}
		\caption{Details of the experimental parameters used in IBM quantum processor IBM QX2 as available on the website \cite{backendQx2}. 
The first row is qubit index q[i] in IBM quantum computer. The second row shows resonance frequencies $\omega^{R}$ of corresponding read-out resonators. The qubit frequencies $\omega$ are given in the  third column. Anhormonicity $\delta$ as provided in the fourth row, is a measure of information leakage out of the computational space. We can compute $\delta$  by taking the difference between two subsequent transition frequencies. The fifth and the sixth rows are qubit-cavity coupling strengths  \textbf{$\chi$} and coupling of the cavity to the environment $\kappa$ for corresponding qubit. Longitudinal relaxation time (T$_{1}$) and Transverse relaxation time (T$_{2}$) are given in the seventh and the last rows, respectively.} \label{exppar1}
	\end{table}

\begin{table}[h]
	
\begin{center}
	\begin{tabular}{|>{\centering}p{2cm}|>{\centering}p{1.5cm}|>{\centering}p{1.5cm}|>{\centering}p{1.5cm}|>{\centering}p{1.5cm}|>{\centering}p{1.5cm}|}
			\hline 
		    Qubit number q[i] & q[0]&q[1]&q[2]&q[3]&q[4] \tabularnewline
			\hline
			$\omega^{R}/2\pi(\operatorname{GHz})$&6.52396&6.48078&6.43875&6.58036&6.52698 \tabularnewline
			\hline
       	$\omega/2\pi(\operatorname{GHz})$&5.2461&5.3025&5.3562&5.4317&5.1824 \tabularnewline
			\hline 
		$\delta/2\pi(\operatorname{MHz})$&-330.1&-329.7&-323.0&-327.9&-332.5 \tabularnewline
			\hline 
		$\textbf{$\chi$}/2\pi(\operatorname{kHz})$&410&512&408&434&458 \tabularnewline 
		\hline 	
		$\operatorname T_{1}(\mu \operatorname s)$ &48.70&39.70&49.70&35.80&56.60 \tabularnewline
		\hline
		$\operatorname T_{2}(\mu \operatorname s)$ &14.00 &34.80&55.00&18.10&31.5 \tabularnewline	
		\hline
		\end{tabular} 
		\end{center}
		\caption{ Details of the experimental parameters used in IBM quantum processor IBM QX4 as available as backend information in \cite{backendQx4}. 
The first row is qubit index q[i] in IBM quantum computer. The second row shows resonance frequencies $\omega^{R}$ of corresponding read-out resonators. The qubit frequencies $\omega$ are given in the  third column. Anharmonicity $\delta$ as provided in the fourth row, is a measure of information leakage out of the computational space. As before, $\delta$ can be calculated by taking the difference between two subsequent transition frequencies. The fifth and the sixth rows are qubit-cavity coupling strengths  \textbf{$\chi$} and coupling of the cavity to the environment $\kappa$ for corresponding qubit. Longitudinal relaxation time (T$_{1}$) and Transverse relaxation time (T$_{2}$) are given in the seventh and the last rows respectively.} \label{exppar2}
	\end{table}
	\end{widetext}
As explained in Sec. \ref{intro}, method for QPT inherently involves QST. Procedure for performing QST on IBM's quantum processors is extensively described in an earlier work of ours \cite{abc}. In the following, we revisit the essential elements of QST.

\begin{enumerate}
	\item Prepare the density matrix which is to be tomographed.
    \item Measure direct observables i.e., perform measurement in computational basis. 
    \item Transform unobservable elements to the directly observable elements by means of a unitary transformation followed by measurement in computational basis.
    \item Apply Hadamard gate to transfer information encoded in $\{+,-\}$ basis to computational basis $\{0,1\}$. Subsequently, perform measurement in $\operatorname{Z}$ (computational) basis for each qubit. 
    \item Apply $S^{\dagger}$ followed by Hadamard $\operatorname{H}$ to transfer information encoded in $\{\ket{r^{+}} = \frac{\ket{0} + i \ket{1}}{\sqrt{2}}, \ket{r^{-}} = \frac{\ket{0}  -i\ket{1}}{\sqrt{2}}\}$ basis to computational basis $\{0,1\}$. Subsequently, measure in $\operatorname{Z}$ basis for each qubit.
    \item Collect probabilities from experimental outcomes and calculate expectation values of all observables constituting density matrix in Pauli basis.
    \item Calculate $\expv{\operatorname{A}}= \sum_{i}p_{i}e_{i}$, where $p_{i}$ is the $i^{th}$ outcome of measurement and $e_{i}$ is the $i^{th}$ eigen value of operator $\operatorname{A}$.
\end{enumerate}

Of course QST requires multiple experiments and the number of experiments to be performed increases with the dimension of the density matrix to be tomographed. Still it's a less complex task in comparison to QPT. This is one of the reasons that QST is reported in a handful of papers on IBM quantum computers \cite{behera2017experimental,behera2018designing,abc} but QPT procedure has not yet been reported in any work. This point will be further clarified below as we would explain the procedure adopted here for performing the experiments required for QPT.

\subsection{Single-qubit QPT}
\begin{figure*}
	\begin{center}
	\includegraphics[scale=0.5]{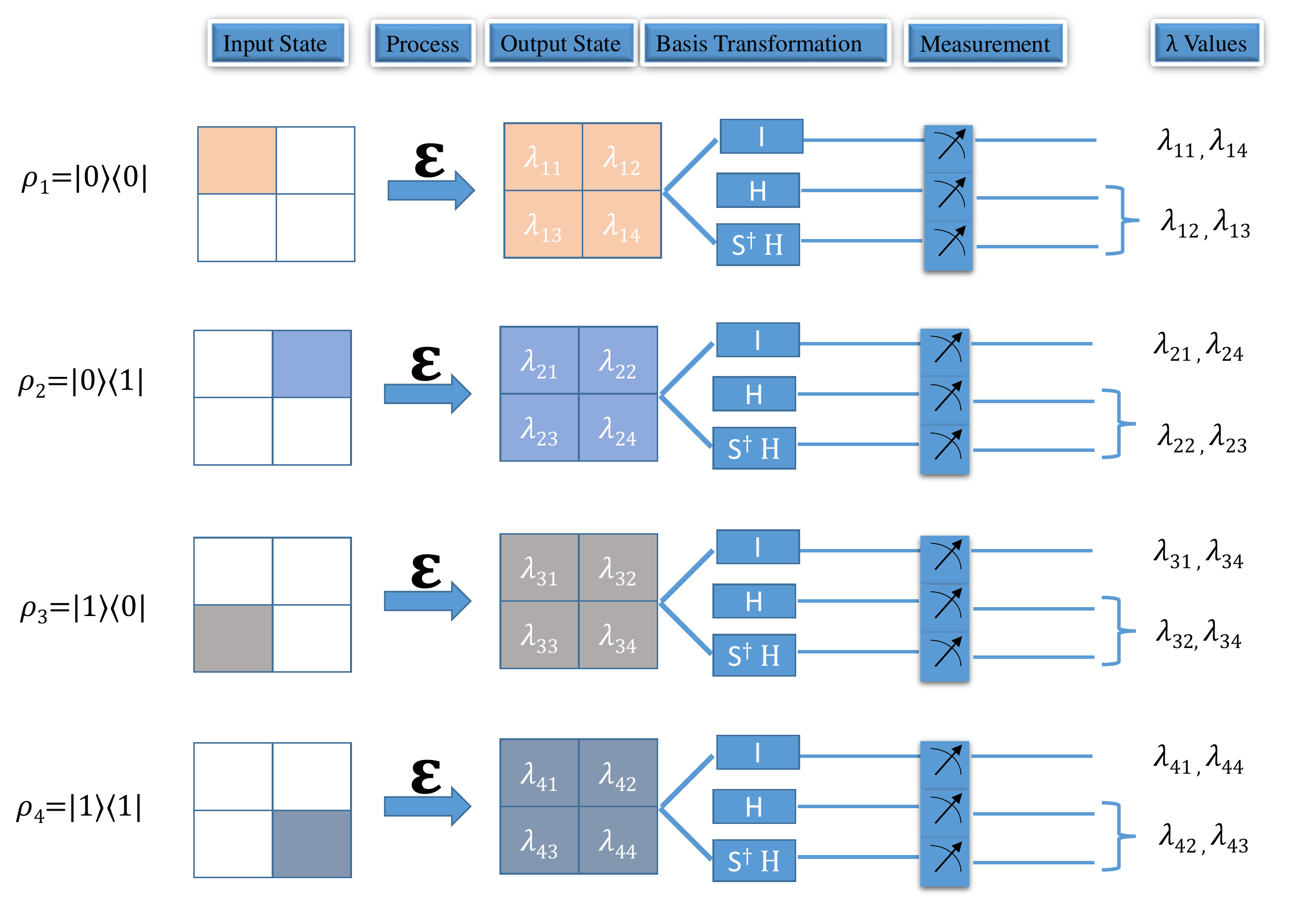}
	\end{center}
	\caption{Scheme for performing single-qubit QPT on IBM's quantum processors. The first column shows input state initialized to basis $\rho_{j}$ followed by application of the quantum operation  $\epsilon(\rho)$ as mentioned in the second column. The Third column contains output states $\rho_{k}$ as a result of application of quantum operation. The fourth and fifth columns are steps of standard procedure for quantum state tomography \cite{ChuangPRL98,cory2000nmr,cory2000nmr1}. The fourth column contains gates/combination of gates used to transfer unobservable non-diagonal elements $\expv{\operatorname{X}}$ and $\expv{\operatorname{Y}}$ into observable diagonal elements while the fifth column shows measurement in $\operatorname{Z}$ basis. Measurement outcomes obtained from three experiments are shown in the last column.}
	\label{expscheme} 
	\end{figure*}

	Following the procedure explained in Sec. \ref{theory}, in what follows, we explain the procedure for QPT adopted here to perform complete characterization of single-qubit gates used in  IBM QX2 and IBM QX4 quantum processors. A schematic diagram illustrating the procedure devised for the single-qubit process tomography for our choice of basis ${\rho_{i}: \rho_{1}=\ket{0}\bra{0}, \rho_{2}=\ket{0}\bra{1}, \rho_{3}=\ket{1}\bra{0}, \rho_{4}=\ket{1}\bra{1}}$ and fixed set of operators $\tilde{E_{m}}= \{\operatorname{\mathbb{I}}, \operatorname{X}, -i\operatorname{Y}, \operatorname{Z}\}$ is shown in Fig. \ref{expscheme}. Execution of single-qubit process tomography on IBM's quantum processors require following steps.

\begin{enumerate}
	\item Initialize density matrix to the basis element $\rho_{1}$ by preparing qubit of interest q[i] to state $\ket{0}$.
    \item Apply quantum process $\epsilon(\rho)$ on initial state $\ket{0}$.
    \item Perform measurement in $\operatorname{Z}$ basis (direct observable) which reveals $\lambda_{11}$ and $\lambda_{14}$.
    \item Apply Hadamard gate to transfer information encoded in $\{+,-\}$ basis to computational basis $\{0,1\}$. Subsequently, measure  in $\operatorname{Z}$ basis.
    \item Apply $S^{\dagger}$ followed by Hadamard $\operatorname{H}$ to transfer information encoded in $\{r^{+},r^{-}\}$  basis to computational basis $\{0,1\}$. Subsequently, measure in $\operatorname{Z}$ basis.
    \item Use above two steps together to obtain $\lambda_{12}$ and $\lambda_{13}$.
	\item Repeat steps 1-5 for all $\rho_{i}$ in order to get all $\lambda_{ij}$.
	\item Compute $\beta^{-1}$ using Eq. \ref{beta} for $\tilde{E_{m}}= \{\operatorname{\mathbb{I}}, \operatorname{X}, -i\operatorname{Y}, \operatorname{Z}\}$.
	\item Finally, calculate vector $\chi$ using Eq. \ref{chi} and thus reconstruct the process matrix $\chi$. 
	\end{enumerate}
	
During experimental implementation of the above procedure for the experimental realization of QPT in the IBM quantum processors one should be careful about a caveat, i.e., $\rho_{2}$ and $\rho_{3}$ are non-Hermitian so it is not possible to prepare them experimentally. In order to overcome this limitation, we have used a concept proposed in \cite{chuangbook}. Specifically, we have performed two experiments as follows: (i) Prepare an initial state $\ket{+}$, apply operation $\epsilon(\rho)$, and tomograph the output state $\epsilon(\ket{+}\bra{+})$. (ii) Do the same experiment  with the initial state $\ket{-}$ and tomograph the output state $\epsilon(\rho)(\ket{-}\bra{-})$. Subsequently, we may compute $\epsilon(\rho_{2})$ and $\epsilon(\rho_{3})$ using the following equations.
 
 \begin{widetext}
 \begin{eqnarray}
 \epsilon(\rho_{2}) = \epsilon(\ket{+}\bra{+}) + i \epsilon(\ket{-}\bra{-}) - (1+i)\frac{(\epsilon(\rho_{1})+ \epsilon(\rho_{4}))}{2},\\
 \epsilon(\rho_{3}) =\epsilon(\ket{+}\bra{+}) - i \epsilon(\ket{-}\bra{-}) - (1-i)\frac{(\epsilon(\rho_{1})+ \epsilon(\rho_{4}))}{2}.
 \end{eqnarray} \label{twoeq}
\end{widetext}	

	\begin{figure}
	\includegraphics[scale=0.25]{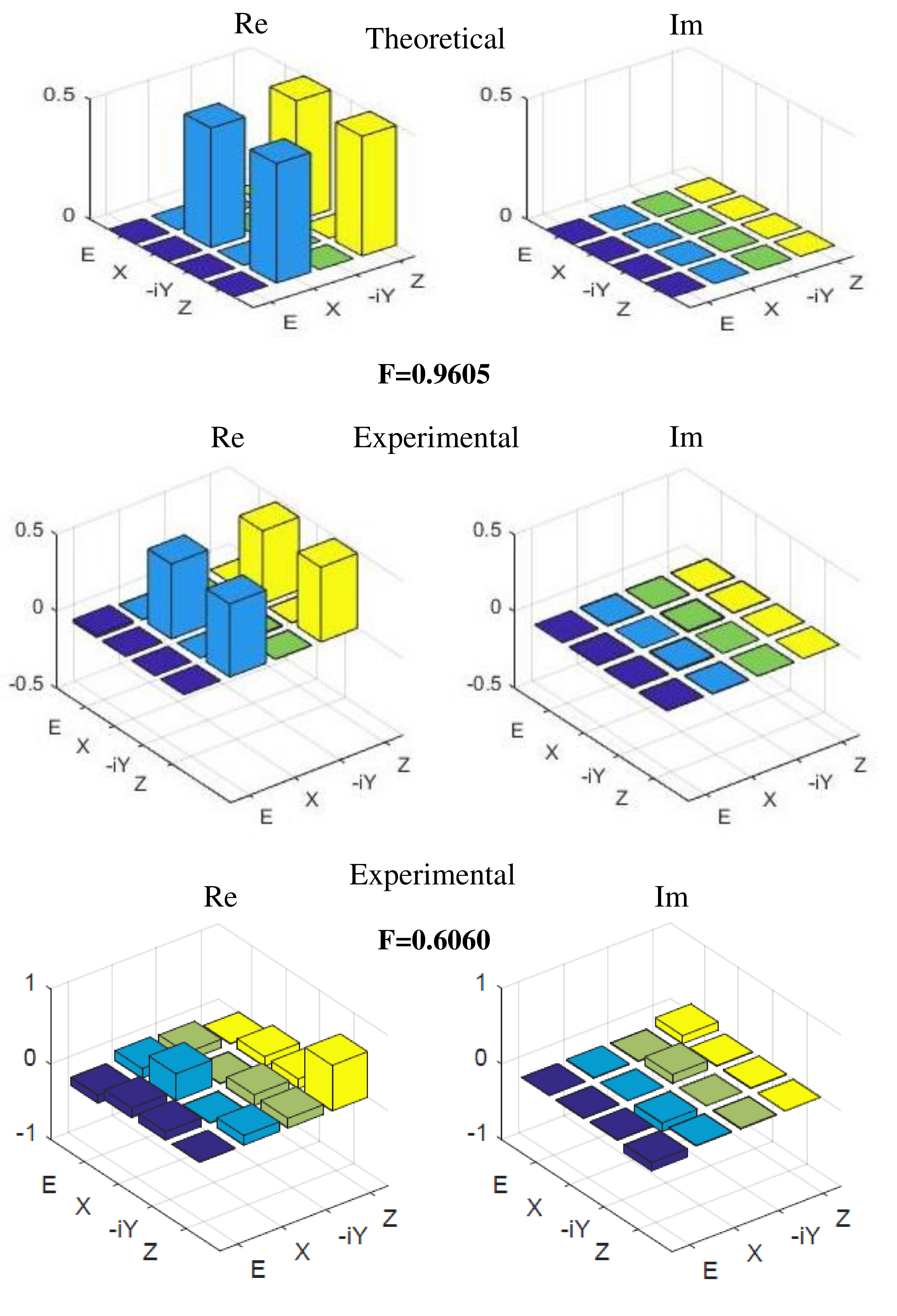}
	\caption{$\chi$ matrices for Hadamard gate obtained from standard quantum process tomography procedure. For the purpose of comparison theoretical $\chi$ matrices (top trace) and experimental  matrices for both IBM QX4 (middle trace) and IBM QX2 (bottom trace) respectively. Left and right columns contain real and imaginary parts of $\chi$ matrix. Fidelity of the Hadamard gates in the two architectures are 0.9650 and 0.6060.}
	\label{chi3hada}
	\end{figure}
	
	\begin{figure}[h]
	\includegraphics[scale=0.35]{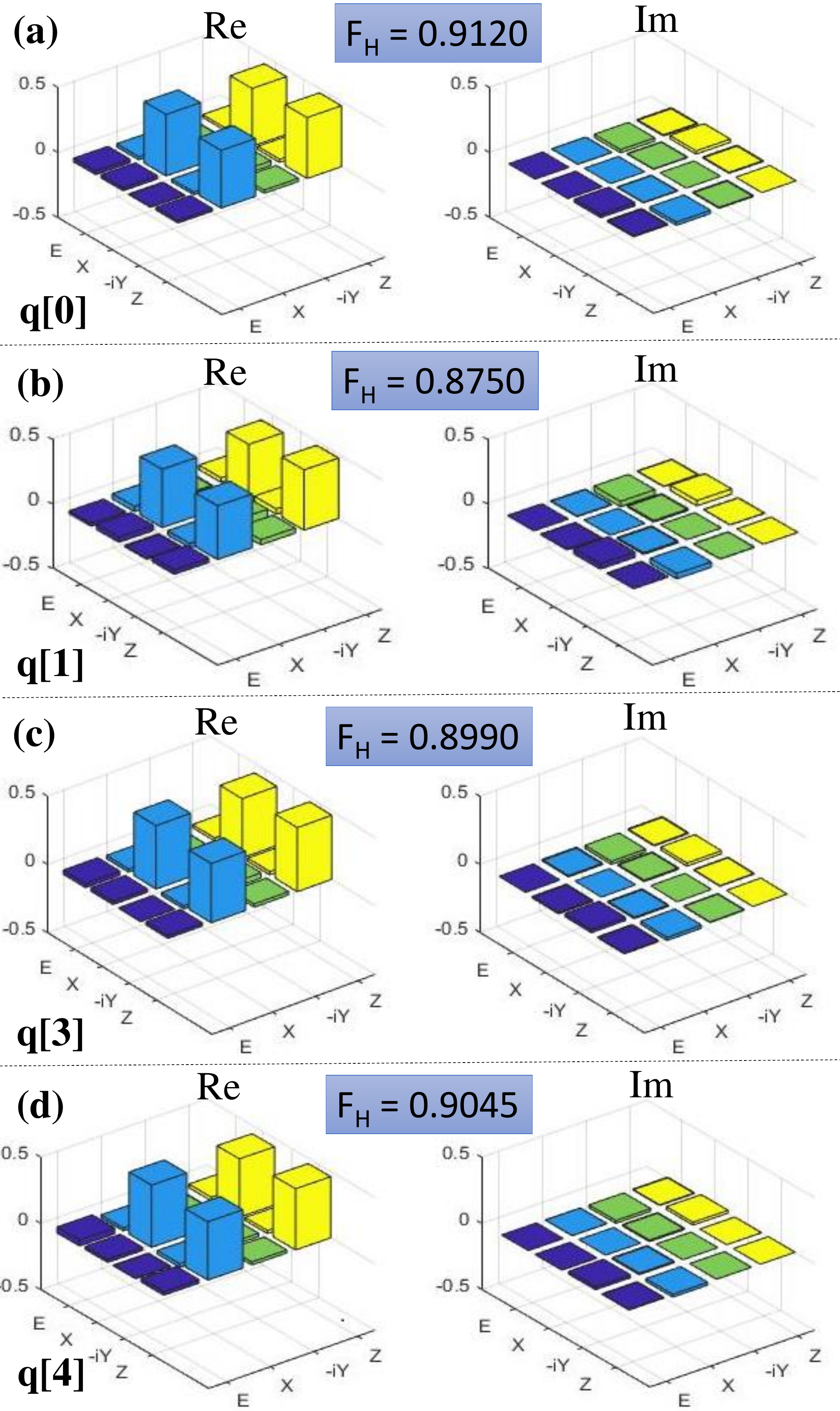}
	\caption{$\chi$ matrices for Hadamard gate applied on different qubits. (a) q[0], (b) q[1], (c) q[3], (d) q[4]. Real and imaginary parts of $\chi$ matrix are arranged as in Fig. \ref{chi3hada}.}
	\label{hadaall}
	\end{figure}
	
	We have used the above scheme to perform QPT of all the single-qubit gates on IBM's quantum processors IBM QX4 (for all the possible positions of the gates) and IBM QX2 (while the gate to be tomographed is placed in the third qubit line only). Results of QPT has been reported in Sec. \ref{results} and summarized in Table \ref{allfidelities} for IBM QX4 processor and in Table \ref{allfidelities1} for IBM QX2 processor. 
		
\begin{widetext}

\begin{table}[h]
\begin{center}
	\begin{tabular}{|>{\centering}p{2cm}|>{\centering}p{1.5cm}|>{\centering}p{1.5cm}|>{\centering}p{1.5cm}|>{\centering}p{1.5cm}|>{\centering}p{1.5cm}|}
		\hline 
		Qubit number q[i] & q[0]&q[1]&q[2]&q[3]&q[4] \tabularnewline
			\hline
		$\operatorname{\mathbb{I}}$&0.9260&0.9073&0.9540&0.9090&0.8872 \tabularnewline
			\hline
       	$\operatorname{X}$&0.9093&0.8850&0.9535&0.8958&0.8855 \tabularnewline
			\hline 
		$\operatorname{Y}$&0.9145&0.8890&0.9515&0.8930&0.8845 \tabularnewline
			\hline 
		$\operatorname{Z}$&0.8222&0.8925&0.9573&0.9042&0.8920 \tabularnewline
		\hline 	
		$\operatorname{H}$ &0.9120&0.8750&0.9605&0.8990&0.9045 \tabularnewline
		\hline
		$\operatorname{T}$ &0.9111&0.9150&0.9611&0.8846&0.8975\tabularnewline	
		\hline
		$\operatorname{T^{\dagger}}$ &0.9049&0.9391&0.9604&0.8688&0.7841\tabularnewline
		\hline
		$\operatorname{S}$&0.8978&0.9147&0.9560&0.8958&0.8978\tabularnewline
		\hline
		$\operatorname{S^{\dagger}}$ &0.9045&0.9445&0.9603&0.8885&0.7948\tabularnewline
		\hline
		\end{tabular} 
		
		\caption{Fidelities of experimental $\chi$ matrices with theoretical $\chi$ matrices for all the single-qubit gates that can be implemented in IBM QX4 quantum processor. The first row is qubit index q[i] as mentioned in IBM quantum computer. Corresponding gates for rows numbers (2-9) are $\colon$ $\operatorname{\mathbb{I}}$, $\operatorname{X}$, $\operatorname{Y}$, $\operatorname{Z}$, $\operatorname{H}$, $\operatorname{T}$, $\operatorname{T}^\dagger$, $\operatorname{S}$, and $\operatorname{S}^{\dagger}$.} \label{allfidelities}
	\end{center}
	\end{table}
\end{widetext}

	\begin{figure*}
	\begin{center}
	\includegraphics[scale=0.5]{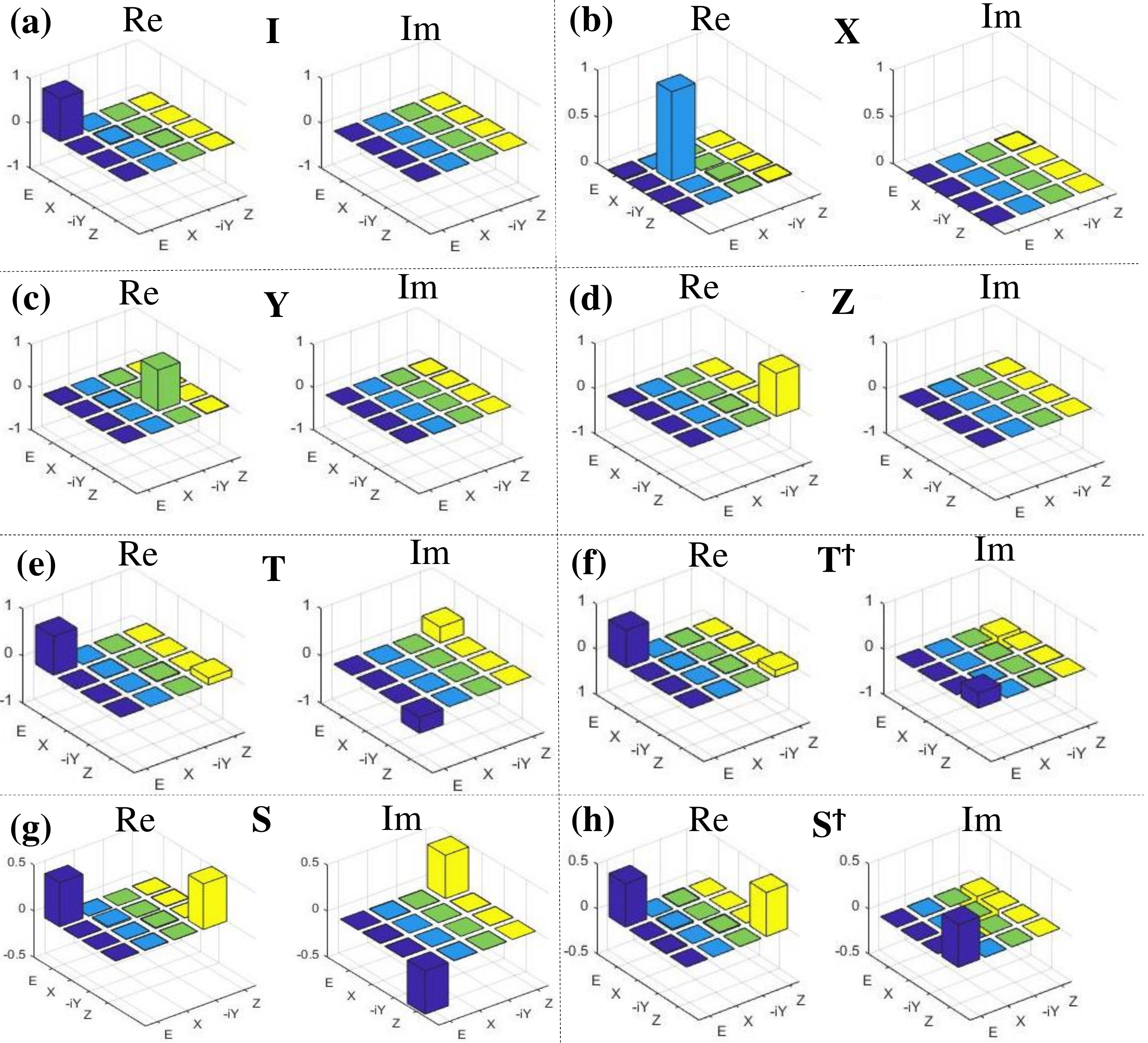}
	\end{center}
	\caption{$\chi$ matrices for various single-qubit gates in IBM's gate library i.e., Clifford$+$T library applied on qubit q[2] of IBM QX4 processor. (a) $\operatorname{\mathbb{I}}$, (b) $\operatorname{X}$, (c) $\operatorname{Y}$, (d) $\operatorname{Z}$, (e) $\operatorname{T}$, (f) $\operatorname{T}^{\dagger}$, (g) $\operatorname{S}$, (f) $ \operatorname{S}^{\dagger}$. Again real and imaginary parts of $\chi$ matrix are arranged as in Fig. \ref{chi3hada}.}
	\label{allgates}
	\end{figure*}
	
	\begin{widetext}
\begin{table*}[t]
\begin{center}
	\begin{tabular}{|>{\centering}p{1.3cm}|>{\centering}p{1.1cm}|>{\centering}p{1.1cm}|>{\centering}p{1.1cm}|>{\centering}p{1.1cm}|>{\centering}p{1.1cm}|>{\centering}p{1.1cm}|>{\centering}p{1.1cm}|>{\centering}p{1.1cm}|>{\centering}p{1.1cm}|}
		\hline 
		Gates &$\operatorname{\mathbb{I}}$&$\operatorname{X}$&$\operatorname{Y}$&$\operatorname{Z}$&$\operatorname{H}$&$\operatorname{T}$&$\operatorname{T}^{\dagger}$&$\operatorname{S}$&$\operatorname{S}^{\dagger}$ \tabularnewline
		\hline                                                                
		Fidelities&0.4855&0.4792&0.9487&0.7190&0.6060&0.9453&0.4746&0.9317&0.9419 \tabularnewline
		\hline
		\end{tabular} 
		\end{center}
		\caption{Fidelities of experimental $\chi$ matrices with theoretical $\chi$ matrices for all the single-qubit gates are obtained by placing the on the third-qubit line in IBM QX2 processor. Corresponding gates for columns (2-9) are $\colon$ $\operatorname{\mathbb{I}}$, $\operatorname{X}$, $\operatorname{Y}$, $\operatorname{Z}$, $\operatorname{H}$, $\operatorname{T}$, $\operatorname{T}^\dagger$, $\operatorname{S}$, and $\operatorname{S}^{\dagger}$.} \label{allfidelities1}
	\end{table*}
		\end{widetext}
\subsection{QPT of C-NOT}
\
In the following, we discuss QPT of C-NOT gates used in the IBM QX4 processor. The only two-qubit gate that can be implemented directly in IBM QX2 and IBM QX4 is CNOT. For our experiments, we have used IBM QX4 version of the processor (now it has been changed to IBM Q5 tenerife). This 5-qubit quantum processor used to allow direct implementation of C-NOT gates in six configurations out of total 20 possible configurations. We have performed QPT of C-NOT gates only for three such positions as that would be enough to establish our point. These configurations are
\begin{enumerate}
\item Qubit q[1] as control and q[0] as target i.e., $C_{1}N_{0}$.
\item Qubit q[2] as control and q[0] as target i.e., $C_{2}N_{0}$.
\item Qubit q[3] as control and q[2] as target i.e., $C_{3}N_{2}$.
\end{enumerate}  

Following the theory of QPT as explained in Sec. \ref{theory} we have computed the relevant $\chi$ matrices. For the purpose of computing $\chi$ matrices,   we choose input basis as 
$\rho_{i}= \{\ket{00}\bra{00}, \ket{00}\bra{01}, \ket{00}\bra{10}, \ket{00}\bra{11}, \ket{01}\bra{00}, \ket{01}\bra{01}\\, \ket{01}\bra{10}, \ket{01}\bra{11}, \ket{10}\bra{00}, \ket{10}\bra{01}, \ket{10}\bra{10}, \ket{10}\bra{11}\\, \ket{11}\bra{00}, \ket{11}\bra{01}, \ket{11}\bra{10}, \ket{11}\bra{11}\}$, for $i=1\colon16$ and fixed set of operators 
$\tilde{\operatorname{E}}= \{\operatorname{I}  \operatorname{I}, \operatorname{I} \operatorname{X},  -i\operatorname{I}\operatorname{Y}, \operatorname{I}  \operatorname{Z}, \operatorname{X}  \operatorname{I}, \operatorname{X} \operatorname{X},   -i\operatorname{X} \operatorname{Y}, \operatorname{X}\operatorname{Z}, -i\operatorname{Y}\operatorname{I}, -i\operatorname{Y}\operatorname{X}\\, -i\operatorname{Y}\operatorname{Y}, -i\operatorname{Y}\operatorname{Z}, \operatorname{Z}  \operatorname{I}, \operatorname{Z}  \operatorname{X},   -i\operatorname{Z}\operatorname{Y}, \operatorname{Z}\operatorname{Z}\}$.  

Further, we followed the steps demonstrated in Sec. \ref{theory} (i.e., to initialize a two-qubit system under consideration into input basis $\rho_{i}$ followed by application of process $\epsilon$ and ultimately performing QST of two-qubit output states $\rho_{j}$) to get all the $\lambda_{ij}$ values. Further, we compute $\beta$ matrix using $\rho_{i}, \rho_{j}, \tilde{\operatorname{E_{m}}}, and \tilde{\operatorname{E_{n}}}$ and ultimately $\chi$ matrix.  
 
 Similar to single-qubit QPT case, also in case of C-NOT gate initialization of non-Hermitian input states i.e., $\rho_{i}-\{\ket{00}\bra{00}, \ket{01}\bra{01}, \ket{10}\bra{10}, \ket{11}\bra{11}\}$ are not allowed which restricts further application of QPT, instead we use appropriate linear combination of elements from the following set of physical states (i.e., $\ket{0p}\bra{0p}, \ket{0m}\bra{0m}, \ket{1p}\bra{1p}, \ket{1m}\bra{1m}, \ket{p0}\bra{p0}\\, \ket{m0}\bra{m0}, \ket{p1}\bra{p1}, \ket{m1}\bra{m1}, \ket{pm}\bra{pm}, \ket{mp}\bra{mp}\\, \ket{pp}\bra{pp}, \ket{mm}\bra{mm}$) in order to realize input states with the help of formula used in \ref{twoeq}.             
	\begin{figure*}
	\begin{center}
	\includegraphics[scale=1]{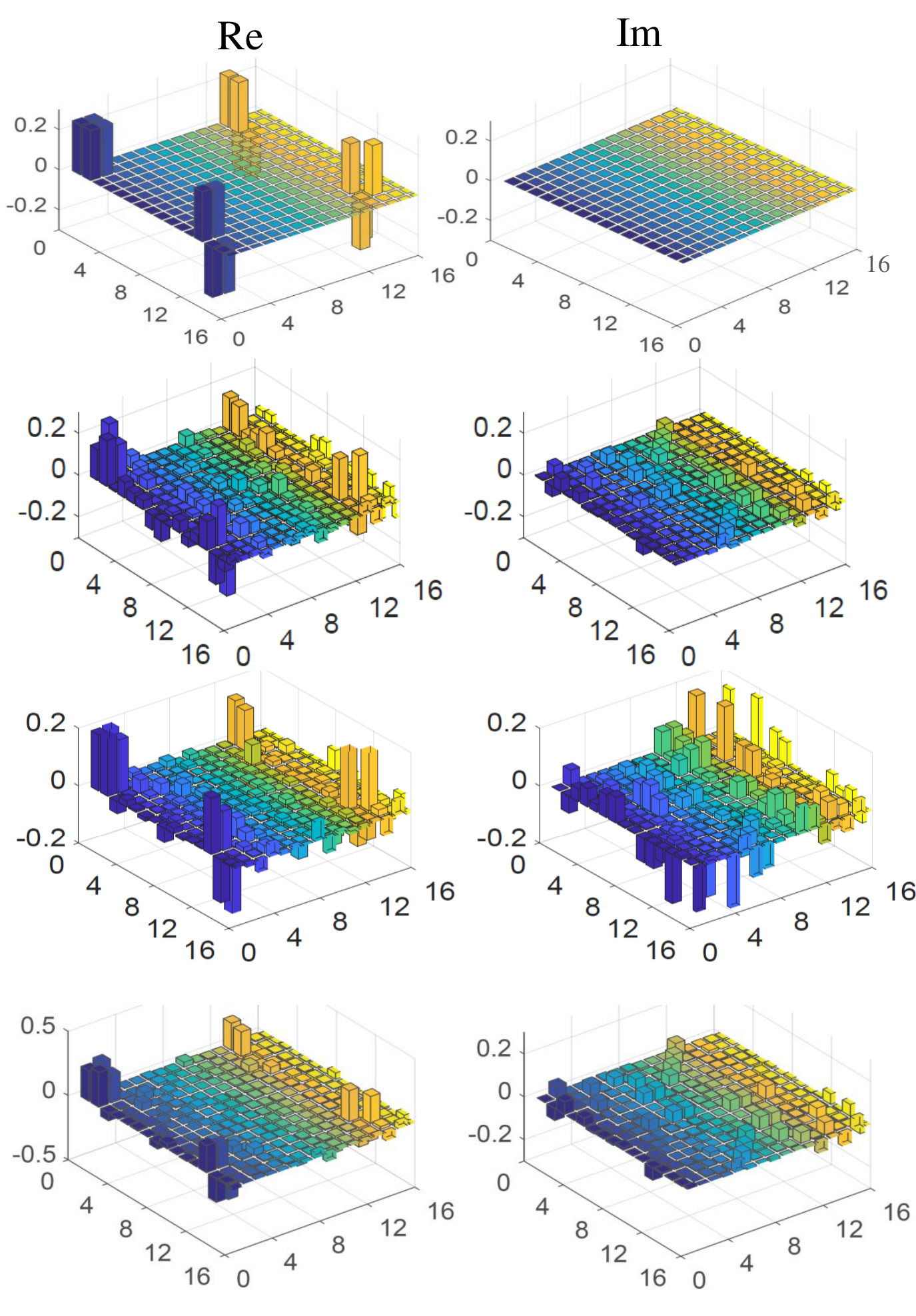}
	\end{center}
          \caption{$\chi$ matrices for various C-NOT gates in IBM's QX4 processor. Theoretical $\chi$ matrices in the top trace, $\chi$ matrices for $C_{1}N_{0}$ gate in the upper middle trace. $\chi$ matrices for $C_{2}N_{0}$ gate in the lower middle trace. $\chi$ matrices for $C_{3}N_{2}$ gate in the bottom trace.}
	\label{cnot}
	\end{figure*}

\color{black}
\section{Results and Discussions} \label{results}
Following the procedure described in the previous section, QPT for various single-qubit gates and C-NOT gates in three configurations that can be realized on IBM's quantum computing architectures have been performed. Representative results for IBM QX4  are illustrated in Fig. \ref{chi3hada}, Fig. \ref{hadaall}, Fig. \ref{allgates}, and Fig. \ref{cnot}. These results can be used to compute fidelity. In fact, fidelity \cite{zhang2014protected} of experimentally obtained $\chi$ matrices with the theoretical ones may be viewed as the gate fidelity  and it has been computed by using the following formula

\begin{equation}
F(	\chi_{th},\chi_{exp}) = \frac{\tr{(\chi_{exp}\chi_{th}^{\dagger})}}{\sqrt{\tr{{\chi_{th}}^{\dagger}\chi_{th}}}\sqrt{\tr{\chi_{exp}^{\dagger}\chi_{exp}}}}.
\end{equation} \label{fidelity}	
Here, $\chi_{th}$ and $\chi_{exp}$ are theoretical and experimental $\chi$ matrices, respectively. For single-qubit case, we perform QPT of all the gates namely, (a) $\operatorname{\mathbb{I}} $, (b) $\operatorname{X}$, (c) $\operatorname{Y}$, (d) $\operatorname{Z}$, (e) $\operatorname{T}$, (f) $\operatorname{T}^{\dagger}$, (g) $\operatorname{S}, (f) \operatorname{S}^{\dagger}$ for all qubit lines for IBM QX4 architecture and for the third qubit line (i.e., q[2]) for IBM QX2. Fidelities for all the single-qubit gates and for all the possible qubit lines for IBM QX4 are given in Table \ref{allfidelities}. Similarly, in Table \ref{allfidelities1}, we provide the results for IBM QX2.  In Fig. \ref{chi3hada}, we show real and imaginary part of $\chi$ matrix for $\operatorname{H}$ gate on third-qubit for both processors. $\chi$ matrices for IBM QX2 with fidelity F$_{\operatorname{H}}$=0.9650 are shown in middle trace and for IBM QX4 with fidelity F$_{\operatorname{H}}$=0.6060 are shown in lower trace. Furthermore, in Fig. \ref{hadaall} we report real and imaginary part of $\chi$ matrices for $\operatorname{H}$ applied on all qubits along with their fidelities. Real and imaginary part of $\chi$ matrices for all the gates applied on third-qubit, for IBM QX2 architecture are shown in Fig. \ref{allgates}.  Results of QPT have been summarized in Table \ref{allfidelities1} for IBM QX2 processor and in Table \ref{allfidelities} for IBM QX4 processor. Comparing the values of fidelity mentioned in these two tables, one can easily conclude that the IBM QX2 would usually perform better than IBM QX4, as IBM QX4's gate fidelity is always higher than that of IBM QX2 (except for the case of $\operatorname{Y}$ gate. Similarly, in case of two-qubit gates we have performed QPT of C-NOT gates $C_{1}N_{0}$, $C_{2}N_{0}$, and $C_{3}N_{2}$. The $\chi$ matrices obtained for these C-NOT gates are shown in  Fig. \ref{cnot}. Top trace contains theoretical $\chi$ matrices, upper middle trace contains experimental $\chi$ matrix of $C_{1}N_{0}$, lower middle trace contains experimental $\chi$ matrix of $C_{2}N_{0}$, and bottom trace contains experimental $\chi$ matrix of $C_{3}N_{2}$. Left column in Fig. \ref{cnot} corresponds to the real part of $\chi$ matrix and right column corresponds to imaginary part of $\chi$ matrix. Fidelity of C-NOT gates  are obtained as $f_{C_{3}N_{2}}= 0.8266$, $f_{C_{1}N_{0}}=  0.7092$, and $f_{C_{2}N_{0}}= 0.6973$, which does not appear to be high. However, to have a feeling of the quality of the gates realized in IBM quantum processors, the comparison  should actually be made with the other technologies. In what follows, we have tried to do so.

We may compare our results with other reported QPT results to develop a feeling about the quality (goodness) of single-qubit gates that can be implemented in IBM's quantum computing architecture. For example, single-qubit gate fidelity, in superconducting qubit based architecture has been reported as 0.98 \cite{lucero2008high}, in nitrogen vacancy center based architecture has been reported as 0.99 \cite{doherty2013nitrogen}, and average fidelity of single-qubit gates in nuclear magnetic resonance architecture has been reported as 0.98 \cite{sspt}. A glance on our results reveals $\operatorname{H}$ gate on third-qubit line has the best gate fidelity with maximum fidelity of 0.97 (cf. Fig.  \ref{chi3hada}). A comparison of IBM's best gate fidelity with gate fidelities achieved in other architectures reveal that the quality of the gates implemented in IBM quantum processors are lower in comparison to other architectures. Similarly, fidelity of C-NOT gate in NMR architecture is `0.99 while in NV center architecture it is 0.98. Although there are several process tomography techniques and which one to use is subject of their best applicability, unfortunately measures for the goodness of gates used in these techniques are not monotone of each other, so a direct comparison of gate fidelities from two techniques (for example as in present case Randomized Bench Marking \cite{chow2009randomized} and standard QPT \cite{chuang97}) is neither feasible \cite{mohseni2008quantum} nor is the aim of the present study, instead we just wish to report bad quality of single-qubit gates as revealed by standard quantum process tomography.

\section{Conclusions} \label{conclusions}
IBM quantum computers allow a user to utilize gates from the Clifford+T gate library with some restrictions on the applicability of C-NOT gates based on the selected architecture. In this work, we have performed QPT for all the quantum gates that can be directly  implemented in IBM QX2 and IBM QX4 architectures. Specifically, QPT has been performed and thus gate fidelity is computed for (i) all the single-qubit gates in all possible positions (i.e., gate fidelity is computed by placing each of the gates in all five qubit lines) in the IBM QX4 architecture, (ii) all the single qubit gates placed individually in the third qubit line in IBM QX2 architecture and (iii) C-NOT gates in IBM QX4 architecture  in three allowed configurations namely, $C_{1}N_{0}$, $C_{2}N_{0}$, $C_{3}N_{2}$. The obtained gate fidelities show that except for the $\operatorname{Y}$ gate, the fidelity of the quantum gates implemented in IBM QX4 is better than the same in IBM QX2. Further, it's observed that the obtained gate fidelity is often lower than the corresponding values obtained earlier using NMR and other technologies. Specially, fidelities of C-NOT gates are found to be really low.  Only $C_{3}N_{2}$ has a reasonably good fidelity.  In brief, the quality of the gates implemented in IBM QX2 and IBM QX4 are not at per with the best results produced in other technologies. This is what has led to lower state fidelity in various cases reported earlier. Further this limitation of the gates has restricted the maximum number gates that can be applied in a particular architecture of IBM quantum computers. This also indicates that the superconductivity-based technology used in IBM quantum computers need to be improved considerably before a scalable quantum computer can be built using this technology. Here comes the most important relevance of the present work- as it provides a method for complete characterization of the process, it may be used to improve the control fields and thus help in the journey of building scalable (or at least relatively bigger) quantum computers.The procedure devised here is quite general.  Keeping these points in mind, we conclude this article with an optimistic view that the present work will not only provide physical insights into the earlier reported results, it will also contribute to the future development of the related technologies.

\textbf{Acknowledgment:}{AP and AS thank Defense Research and Development Organization (DRDO), India for the support provided through the project number ERIPR/ER/1403163/M/01/1603. AS also thanks University of Science and Technology of China, Hefei, People republic of China for their support.}
\bibliographystyle{apsrev4-1}
\bibliography{test6}

\begin{thebibliography}{47}%
\makeatletter
\providecommand \@ifxundefined [1]{%
 \@ifx{#1\undefined}
}%
\providecommand \@ifnum [1]{%
 \ifnum #1\expandafter \@firstoftwo
 \else \expandafter \@secondoftwo
 \fi
}%
\providecommand \@ifx [1]{%
 \ifx #1\expandafter \@firstoftwo
 \else \expandafter \@secondoftwo
 \fi
}%
\providecommand \natexlab [1]{#1}%
\providecommand \enquote  [1]{``#1''}%
\providecommand \bibnamefont  [1]{#1}%
\providecommand \bibfnamefont [1]{#1}%
\providecommand \citenamefont [1]{#1}%
\providecommand \href@noop [0]{\@secondoftwo}%
\providecommand \href [0]{\begingroup \@sanitize@url \@href}%
\providecommand \@href[1]{\@@startlink{#1}\@@href}%
\providecommand \@@href[1]{\endgroup#1\@@endlink}%
\providecommand \@sanitize@url [0]{\catcode `\\12\catcode `\$12\catcode
  `\&12\catcode `\#12\catcode `\^12\catcode `\_12\catcode `\%12\relax}%
\providecommand \@@startlink[1]{}%
\providecommand \@@endlink[0]{}%
\providecommand \url  [0]{\begingroup\@sanitize@url \@url }%
\providecommand \@url [1]{\endgroup\@href {#1}{\urlprefix }}%
\providecommand \urlprefix  [0]{URL }%
\providecommand \Eprint [0]{\href }%
\providecommand \doibase [0]{http://dx.doi.org/}%
\providecommand \selectlanguage [0]{\@gobble}%
\providecommand \bibinfo  [0]{\@secondoftwo}%
\providecommand \bibfield  [0]{\@secondoftwo}%
\providecommand \translation [1]{[#1]}%
\providecommand \BibitemOpen [0]{}%
\providecommand \bibitemStop [0]{}%
\providecommand \bibitemNoStop [0]{.\EOS\space}%
\providecommand \EOS [0]{\spacefactor3000\relax}%
\providecommand \BibitemShut  [1]{\csname bibitem#1\endcsname}%
\let\auto@bib@innerbib\@empty
\bibitem [{\citenamefont {Chuang}\ and\ \citenamefont
  {Nielsen}(1997)}]{chuang97}%
  \BibitemOpen
  \bibfield  {author} {\bibinfo {author} {\bibfnamefont {I.~L.}\ \bibnamefont
  {Chuang}}\ and\ \bibinfo {author} {\bibfnamefont {M.~A.}\ \bibnamefont
  {Nielsen}},\ }\href@noop {} {\bibfield  {journal} {\bibinfo  {journal} {J.
  Mod. Opt.}\ }\textbf {\bibinfo {volume} {44}},\ \bibinfo {pages} {2455}
  (\bibinfo {year} {1997})}\BibitemShut {NoStop}%
\bibitem [{\citenamefont {Poyatos}\ \emph {et~al.}(1997)\citenamefont
  {Poyatos}, \citenamefont {Cirac},\ and\ \citenamefont {Zoller}}]{zollar97}%
  \BibitemOpen
  \bibfield  {author} {\bibinfo {author} {\bibfnamefont {J.}~\bibnamefont
  {Poyatos}}, \bibinfo {author} {\bibfnamefont {J.~I.}\ \bibnamefont {Cirac}},
  \ and\ \bibinfo {author} {\bibfnamefont {P.}~\bibnamefont {Zoller}},\
  }\href@noop {} {\bibfield  {journal} {\bibinfo  {journal} {Phys. Rev. Lett.}\
  }\textbf {\bibinfo {volume} {78}},\ \bibinfo {pages} {390} (\bibinfo {year}
  {1997})}\BibitemShut {NoStop}%
\bibitem [{\citenamefont {I.~L.~Chuang}\ and\ \citenamefont
  {Leung}(1998)}]{ChuangPRL98}%
  \BibitemOpen
  \bibfield  {author} {\bibinfo {author} {\bibfnamefont {M.~G.~K.}\
  \bibnamefont {I.~L.~Chuang}, \bibfnamefont {N.~Gershenfeld}}\ and\ \bibinfo
  {author} {\bibfnamefont {D.~W.}\ \bibnamefont {Leung}},\ }\href@noop {}
  {\bibfield  {journal} {\bibinfo  {journal} {Proc. R. Soc. Lond.,Ser A}\
  }\textbf {\bibinfo {volume} {454}},\ \bibinfo {pages} {447} (\bibinfo {year}
  {1998})}\BibitemShut {NoStop}%
\bibitem [{\citenamefont {Childs}\ \emph {et~al.}(2001)\citenamefont {Childs},
  \citenamefont {Chuang},\ and\ \citenamefont {Leung}}]{ChuangPRA2001}%
  \BibitemOpen
  \bibfield  {author} {\bibinfo {author} {\bibfnamefont {A.~M.}\ \bibnamefont
  {Childs}}, \bibinfo {author} {\bibfnamefont {I.~L.}\ \bibnamefont {Chuang}},
  \ and\ \bibinfo {author} {\bibfnamefont {D.~W.}\ \bibnamefont {Leung}},\
  }\href@noop {} {\bibfield  {journal} {\bibinfo  {journal} {Phys. Rev. A}\
  }\textbf {\bibinfo {volume} {64}},\ \bibinfo {pages} {012314} (\bibinfo
  {year} {2001})}\BibitemShut {NoStop}%
\bibitem [{\citenamefont {Weinstein}\ \emph {et~al.}(2004)\citenamefont
  {Weinstein}, \citenamefont {Havel}, \citenamefont {Emerson}, \citenamefont
  {Boulant}, \citenamefont {Saraceno}, \citenamefont {Lloyd},\ and\
  \citenamefont {Cory}}]{QPTofQFT}%
  \BibitemOpen
  \bibfield  {author} {\bibinfo {author} {\bibfnamefont {Y.~S.}\ \bibnamefont
  {Weinstein}}, \bibinfo {author} {\bibfnamefont {T.~F.}\ \bibnamefont
  {Havel}}, \bibinfo {author} {\bibfnamefont {J.}~\bibnamefont {Emerson}},
  \bibinfo {author} {\bibfnamefont {N.}~\bibnamefont {Boulant}}, \bibinfo
  {author} {\bibfnamefont {M.}~\bibnamefont {Saraceno}}, \bibinfo {author}
  {\bibfnamefont {S.}~\bibnamefont {Lloyd}}, \ and\ \bibinfo {author}
  {\bibfnamefont {D.~G.}\ \bibnamefont {Cory}},\ }\href@noop {} {\bibfield
  {journal} {\bibinfo  {journal} {J. Chem. Phys.}\ }\textbf {\bibinfo {volume}
  {121}},\ \bibinfo {pages} {6117} (\bibinfo {year} {2004})}\BibitemShut
  {NoStop}%
\bibitem [{\citenamefont {De~Martini}\ \emph {et~al.}(2003)\citenamefont
  {De~Martini}, \citenamefont {Mazzei}, \citenamefont {Ricci},\ and\
  \citenamefont {D'Ariano}}]{EAPT1Exp}%
  \BibitemOpen
  \bibfield  {author} {\bibinfo {author} {\bibfnamefont {F.}~\bibnamefont
  {De~Martini}}, \bibinfo {author} {\bibfnamefont {A.}~\bibnamefont {Mazzei}},
  \bibinfo {author} {\bibfnamefont {M.}~\bibnamefont {Ricci}}, \ and\ \bibinfo
  {author} {\bibfnamefont {G.~M.}\ \bibnamefont {D'Ariano}},\ }\href@noop {}
  {\bibfield  {journal} {\bibinfo  {journal} {Phys. Rev. A}\ }\textbf {\bibinfo
  {volume} {67}},\ \bibinfo {pages} {062307} (\bibinfo {year}
  {2003})}\BibitemShut {NoStop}%
\bibitem [{\citenamefont {Altepeter}\ \emph {et~al.}(2003)\citenamefont
  {Altepeter}, \citenamefont {Branning}, \citenamefont {Jeffrey}, \citenamefont
  {Wei}, \citenamefont {Kwiat}, \citenamefont {Thew}, \citenamefont {O'Brien},
  \citenamefont {Nielsen},\ and\ \citenamefont {White}}]{altepeter}%
  \BibitemOpen
  \bibfield  {author} {\bibinfo {author} {\bibfnamefont {J.~B.}\ \bibnamefont
  {Altepeter}}, \bibinfo {author} {\bibfnamefont {D.}~\bibnamefont {Branning}},
  \bibinfo {author} {\bibfnamefont {E.}~\bibnamefont {Jeffrey}}, \bibinfo
  {author} {\bibfnamefont {T.~C.}\ \bibnamefont {Wei}}, \bibinfo {author}
  {\bibfnamefont {P.~G.}\ \bibnamefont {Kwiat}}, \bibinfo {author}
  {\bibfnamefont {R.~T.}\ \bibnamefont {Thew}}, \bibinfo {author}
  {\bibfnamefont {J.~L.}\ \bibnamefont {O'Brien}}, \bibinfo {author}
  {\bibfnamefont {M.~A.}\ \bibnamefont {Nielsen}}, \ and\ \bibinfo {author}
  {\bibfnamefont {A.~G.}\ \bibnamefont {White}},\ }\href@noop {} {\bibfield
  {journal} {\bibinfo  {journal} {Phys. Rev. Lett.}\ }\textbf {\bibinfo
  {volume} {90}},\ \bibinfo {pages} {193601} (\bibinfo {year}
  {2003})}\BibitemShut {NoStop}%
\bibitem [{\citenamefont {O'Brien}\ \emph {et~al.}(2004)\citenamefont
  {O'Brien}, \citenamefont {Pryde}, \citenamefont {Gilchrist}, \citenamefont
  {James}, \citenamefont {Langford}, \citenamefont {Ralph},\ and\ \citenamefont
  {White}}]{obrian}%
  \BibitemOpen
  \bibfield  {author} {\bibinfo {author} {\bibfnamefont {J.~L.}\ \bibnamefont
  {O'Brien}}, \bibinfo {author} {\bibfnamefont {G.~J.}\ \bibnamefont {Pryde}},
  \bibinfo {author} {\bibfnamefont {A.}~\bibnamefont {Gilchrist}}, \bibinfo
  {author} {\bibfnamefont {D.~F.~V.}\ \bibnamefont {James}}, \bibinfo {author}
  {\bibfnamefont {N.~K.}\ \bibnamefont {Langford}}, \bibinfo {author}
  {\bibfnamefont {T.~C.}\ \bibnamefont {Ralph}}, \ and\ \bibinfo {author}
  {\bibfnamefont {A.~G.}\ \bibnamefont {White}},\ }\href@noop {} {\bibfield
  {journal} {\bibinfo  {journal} {Phys. Rev. Lett.}\ }\textbf {\bibinfo
  {volume} {93}},\ \bibinfo {pages} {080502} (\bibinfo {year}
  {2004})}\BibitemShut {NoStop}%
\bibitem [{\citenamefont {Mitchell}\ \emph {et~al.}(2003)\citenamefont
  {Mitchell}, \citenamefont {Ellenor}, \citenamefont {Schneider},\ and\
  \citenamefont {Steinberg}}]{bellstatefilter}%
  \BibitemOpen
  \bibfield  {author} {\bibinfo {author} {\bibfnamefont {M.~W.}\ \bibnamefont
  {Mitchell}}, \bibinfo {author} {\bibfnamefont {C.~W.}\ \bibnamefont
  {Ellenor}}, \bibinfo {author} {\bibfnamefont {S.}~\bibnamefont {Schneider}},
  \ and\ \bibinfo {author} {\bibfnamefont {A.~M.}\ \bibnamefont {Steinberg}},\
  }\href@noop {} {\bibfield  {journal} {\bibinfo  {journal} {Phys. Rev. Lett.}\
  }\textbf {\bibinfo {volume} {91}},\ \bibinfo {pages} {120402} (\bibinfo
  {year} {2003})}\BibitemShut {NoStop}%
\bibitem [{\citenamefont {Riebe}\ \emph {et~al.}(2006)\citenamefont {Riebe},
  \citenamefont {Kim}, \citenamefont {Schindler}, \citenamefont {Monz},
  \citenamefont {Schmidt}, \citenamefont {K\"orber}, \citenamefont {H\"ansel},
  \citenamefont {H\"affner}, \citenamefont {Roos},\ and\ \citenamefont
  {Blatt}}]{qptITprl2006}%
  \BibitemOpen
  \bibfield  {author} {\bibinfo {author} {\bibfnamefont {M.}~\bibnamefont
  {Riebe}}, \bibinfo {author} {\bibfnamefont {K.}~\bibnamefont {Kim}}, \bibinfo
  {author} {\bibfnamefont {P.}~\bibnamefont {Schindler}}, \bibinfo {author}
  {\bibfnamefont {T.}~\bibnamefont {Monz}}, \bibinfo {author} {\bibfnamefont
  {P.~O.}\ \bibnamefont {Schmidt}}, \bibinfo {author} {\bibfnamefont {T.~K.}\
  \bibnamefont {K\"orber}}, \bibinfo {author} {\bibfnamefont {W.}~\bibnamefont
  {H\"ansel}}, \bibinfo {author} {\bibfnamefont {H.}~\bibnamefont {H\"affner}},
  \bibinfo {author} {\bibfnamefont {C.~F.}\ \bibnamefont {Roos}}, \ and\
  \bibinfo {author} {\bibfnamefont {R.}~\bibnamefont {Blatt}},\ }\href@noop {}
  {\bibfield  {journal} {\bibinfo  {journal} {Phys. Rev. Lett.}\ }\textbf
  {\bibinfo {volume} {97}},\ \bibinfo {pages} {220407} (\bibinfo {year}
  {2006})}\BibitemShut {NoStop}%
\bibitem [{\citenamefont {Hanneke}\ \emph {et~al.}(2010)\citenamefont
  {Hanneke}, \citenamefont {Home}, \citenamefont {Jost}, \citenamefont {Amini},
  \citenamefont {Leibfried},\ and\ \citenamefont
  {Wineland}}]{QPT_IonTrapNature2010}%
  \BibitemOpen
  \bibfield  {author} {\bibinfo {author} {\bibfnamefont {D.}~\bibnamefont
  {Hanneke}}, \bibinfo {author} {\bibfnamefont {J.~P.}\ \bibnamefont {Home}},
  \bibinfo {author} {\bibfnamefont {J.~D.}\ \bibnamefont {Jost}}, \bibinfo
  {author} {\bibfnamefont {J.~M.}\ \bibnamefont {Amini}}, \bibinfo {author}
  {\bibfnamefont {D.}~\bibnamefont {Leibfried}}, \ and\ \bibinfo {author}
  {\bibfnamefont {D.~J.}\ \bibnamefont {Wineland}},\ }\href@noop {} {\bibfield
  {journal} {\bibinfo  {journal} {Nature Phys.}\ }\textbf {\bibinfo {volume}
  {6}},\ \bibinfo {pages} {13} (\bibinfo {year} {2010})}\BibitemShut {NoStop}%
\bibitem [{\citenamefont {Neeley}\ \emph {et~al.}(2008)\citenamefont {Neeley},
  \citenamefont {Ansmann}, \citenamefont {Bialczak}, \citenamefont {Hofheinz},
  \citenamefont {Katz}, \citenamefont {Lucero}, \citenamefont {O/'Connell},
  \citenamefont {Wang}, \citenamefont {Cleland},\ and\ \citenamefont
  {Martinis}}]{QPT_SQUID}%
  \BibitemOpen
  \bibfield  {author} {\bibinfo {author} {\bibfnamefont {M.}~\bibnamefont
  {Neeley}}, \bibinfo {author} {\bibfnamefont {M.}~\bibnamefont {Ansmann}},
  \bibinfo {author} {\bibfnamefont {R.~C.}\ \bibnamefont {Bialczak}}, \bibinfo
  {author} {\bibfnamefont {M.}~\bibnamefont {Hofheinz}}, \bibinfo {author}
  {\bibfnamefont {N.}~\bibnamefont {Katz}}, \bibinfo {author} {\bibfnamefont
  {E.}~\bibnamefont {Lucero}}, \bibinfo {author} {\bibfnamefont
  {A.}~\bibnamefont {O/'Connell}}, \bibinfo {author} {\bibfnamefont
  {H.}~\bibnamefont {Wang}}, \bibinfo {author} {\bibfnamefont {A.~N.}\
  \bibnamefont {Cleland}}, \ and\ \bibinfo {author} {\bibfnamefont {J.~M.}\
  \bibnamefont {Martinis}},\ }\href@noop {} {\bibfield  {journal} {\bibinfo
  {journal} {Nature Phys.}\ }\textbf {\bibinfo {volume} {4}},\ \bibinfo {pages}
  {523} (\bibinfo {year} {2008})}\BibitemShut {NoStop}%
\bibitem [{\citenamefont {Chow}\ \emph
  {et~al.}(2009{\natexlab{a}})\citenamefont {Chow}, \citenamefont {Gambetta},
  \citenamefont {Tornberg}, \citenamefont {Koch}, \citenamefont {Bishop},
  \citenamefont {Houck}, \citenamefont {Johnson}, \citenamefont {Frunzio},
  \citenamefont {Girvin},\ and\ \citenamefont {Schoelkopf}}]{SQUID2009}%
  \BibitemOpen
  \bibfield  {author} {\bibinfo {author} {\bibfnamefont {J.~M.}\ \bibnamefont
  {Chow}}, \bibinfo {author} {\bibfnamefont {J.~M.}\ \bibnamefont {Gambetta}},
  \bibinfo {author} {\bibfnamefont {L.}~\bibnamefont {Tornberg}}, \bibinfo
  {author} {\bibfnamefont {J.}~\bibnamefont {Koch}}, \bibinfo {author}
  {\bibfnamefont {L.~S.}\ \bibnamefont {Bishop}}, \bibinfo {author}
  {\bibfnamefont {A.~A.}\ \bibnamefont {Houck}}, \bibinfo {author}
  {\bibfnamefont {B.~R.}\ \bibnamefont {Johnson}}, \bibinfo {author}
  {\bibfnamefont {L.}~\bibnamefont {Frunzio}}, \bibinfo {author} {\bibfnamefont
  {S.~M.}\ \bibnamefont {Girvin}}, \ and\ \bibinfo {author} {\bibfnamefont
  {R.~J.}\ \bibnamefont {Schoelkopf}},\ }\href@noop {} {\bibfield  {journal}
  {\bibinfo  {journal} {Phys. Rev. Lett.}\ }\textbf {\bibinfo {volume} {102}},\
  \bibinfo {pages} {090502} (\bibinfo {year} {2009}{\natexlab{a}})}\BibitemShut
  {NoStop}%
\bibitem [{\citenamefont {Bialczak}\ \emph {et~al.}(2010)\citenamefont
  {Bialczak}, \citenamefont {Ansmann}, \citenamefont {Hofheinz}, \citenamefont
  {Lucero}, \citenamefont {Neeley}, \citenamefont {O/'Connell}, \citenamefont
  {Sank}, \citenamefont {Wang}, \citenamefont {Wenner}, \citenamefont
  {Steffen},\ and\ \citenamefont {Cleland}}]{MartiniSQUID2010}%
  \BibitemOpen
  \bibfield  {author} {\bibinfo {author} {\bibfnamefont {R.~C.}\ \bibnamefont
  {Bialczak}}, \bibinfo {author} {\bibfnamefont {M.}~\bibnamefont {Ansmann}},
  \bibinfo {author} {\bibfnamefont {M.}~\bibnamefont {Hofheinz}}, \bibinfo
  {author} {\bibfnamefont {E.}~\bibnamefont {Lucero}}, \bibinfo {author}
  {\bibfnamefont {M.}~\bibnamefont {Neeley}}, \bibinfo {author} {\bibfnamefont
  {A.~D.}\ \bibnamefont {O/'Connell}}, \bibinfo {author} {\bibfnamefont
  {D.}~\bibnamefont {Sank}}, \bibinfo {author} {\bibfnamefont {H.}~\bibnamefont
  {Wang}}, \bibinfo {author} {\bibfnamefont {J.}~\bibnamefont {Wenner}},
  \bibinfo {author} {\bibfnamefont {M.}~\bibnamefont {Steffen}}, \ and\
  \bibinfo {author} {\bibfnamefont {J.~M.}\ \bibnamefont {Cleland},
  \bibfnamefont {A.~N.and~Martinis}},\ }\href@noop {} {\bibfield  {journal}
  {\bibinfo  {journal} {Nature Phys.}\ }\textbf {\bibinfo {volume} {6}},\
  \bibinfo {pages} {409} (\bibinfo {year} {2010})}\BibitemShut {NoStop}%
\bibitem [{\citenamefont {Yamamoto}\ \emph {et~al.}(2010)\citenamefont
  {Yamamoto}, \citenamefont {Neeley}, \citenamefont {Lucero}, \citenamefont
  {Bialczak}, \citenamefont {Kelly}, \citenamefont {Lenander}, \citenamefont
  {Mariantoni}, \citenamefont {O'Connell}, \citenamefont {Sank}, \citenamefont
  {Wang}, \citenamefont {Weides}, \citenamefont {Wenner}, \citenamefont {Yin},
  \citenamefont {Cleland},\ and\ \citenamefont {Martinis}}]{QPT2spSQUID}%
  \BibitemOpen
  \bibfield  {author} {\bibinfo {author} {\bibfnamefont {T.}~\bibnamefont
  {Yamamoto}}, \bibinfo {author} {\bibfnamefont {M.}~\bibnamefont {Neeley}},
  \bibinfo {author} {\bibfnamefont {E.}~\bibnamefont {Lucero}}, \bibinfo
  {author} {\bibfnamefont {R.~C.}\ \bibnamefont {Bialczak}}, \bibinfo {author}
  {\bibfnamefont {J.}~\bibnamefont {Kelly}}, \bibinfo {author} {\bibfnamefont
  {M.}~\bibnamefont {Lenander}}, \bibinfo {author} {\bibfnamefont
  {M.}~\bibnamefont {Mariantoni}}, \bibinfo {author} {\bibfnamefont {A.~D.}\
  \bibnamefont {O'Connell}}, \bibinfo {author} {\bibfnamefont {D.}~\bibnamefont
  {Sank}}, \bibinfo {author} {\bibfnamefont {H.}~\bibnamefont {Wang}}, \bibinfo
  {author} {\bibfnamefont {M.}~\bibnamefont {Weides}}, \bibinfo {author}
  {\bibfnamefont {J.}~\bibnamefont {Wenner}}, \bibinfo {author} {\bibfnamefont
  {Y.}~\bibnamefont {Yin}}, \bibinfo {author} {\bibfnamefont {A.~N.}\
  \bibnamefont {Cleland}}, \ and\ \bibinfo {author} {\bibfnamefont {J.~M.}\
  \bibnamefont {Martinis}},\ }\href@noop {} {\bibfield  {journal} {\bibinfo
  {journal} {Phys. Rev. B}\ }\textbf {\bibinfo {volume} {82}},\ \bibinfo
  {pages} {184515} (\bibinfo {year} {2010})}\BibitemShut {NoStop}%
\bibitem [{\citenamefont {Chow}\ \emph {et~al.}(2011)\citenamefont {Chow},
  \citenamefont {C\'orcoles}, \citenamefont {Gambetta}, \citenamefont
  {Rigetti}, \citenamefont {Johnson}, \citenamefont {Smolin}, \citenamefont
  {Rozen}, \citenamefont {Keefe}, \citenamefont {Rothwell}, \citenamefont
  {Ketchen},\ and\ \citenamefont {Steffen}}]{QPT_SQUIDChow2011}%
  \BibitemOpen
  \bibfield  {author} {\bibinfo {author} {\bibfnamefont {J.~M.}\ \bibnamefont
  {Chow}}, \bibinfo {author} {\bibfnamefont {A.~D.}\ \bibnamefont
  {C\'orcoles}}, \bibinfo {author} {\bibfnamefont {J.~M.}\ \bibnamefont
  {Gambetta}}, \bibinfo {author} {\bibfnamefont {C.}~\bibnamefont {Rigetti}},
  \bibinfo {author} {\bibfnamefont {B.~R.}\ \bibnamefont {Johnson}}, \bibinfo
  {author} {\bibfnamefont {J.~A.}\ \bibnamefont {Smolin}}, \bibinfo {author}
  {\bibfnamefont {J.~R.}\ \bibnamefont {Rozen}}, \bibinfo {author}
  {\bibfnamefont {G.~A.}\ \bibnamefont {Keefe}}, \bibinfo {author}
  {\bibfnamefont {M.~B.}\ \bibnamefont {Rothwell}}, \bibinfo {author}
  {\bibfnamefont {M.~B.}\ \bibnamefont {Ketchen}}, \ and\ \bibinfo {author}
  {\bibfnamefont {M.}~\bibnamefont {Steffen}},\ }\href@noop {} {\bibfield
  {journal} {\bibinfo  {journal} {Phys. Rev. Lett.}\ }\textbf {\bibinfo
  {volume} {107}},\ \bibinfo {pages} {080502} (\bibinfo {year}
  {2011})}\BibitemShut {NoStop}%
\bibitem [{\citenamefont {Dewes}\ \emph {et~al.}(2012)\citenamefont {Dewes},
  \citenamefont {Ong}, \citenamefont {Schmitt}, \citenamefont {Lauro},
  \citenamefont {Boulant}, \citenamefont {Bertet}, \citenamefont {Vion},\ and\
  \citenamefont {Esteve}}]{SQUID_Dewes2012}%
  \BibitemOpen
  \bibfield  {author} {\bibinfo {author} {\bibfnamefont {A.}~\bibnamefont
  {Dewes}}, \bibinfo {author} {\bibfnamefont {F.~R.}\ \bibnamefont {Ong}},
  \bibinfo {author} {\bibfnamefont {V.}~\bibnamefont {Schmitt}}, \bibinfo
  {author} {\bibfnamefont {R.}~\bibnamefont {Lauro}}, \bibinfo {author}
  {\bibfnamefont {N.}~\bibnamefont {Boulant}}, \bibinfo {author} {\bibfnamefont
  {P.}~\bibnamefont {Bertet}}, \bibinfo {author} {\bibfnamefont
  {D.}~\bibnamefont {Vion}}, \ and\ \bibinfo {author} {\bibfnamefont
  {D.}~\bibnamefont {Esteve}},\ }\href@noop {} {\bibfield  {journal} {\bibinfo
  {journal} {Phys. Rev. Lett.}\ }\textbf {\bibinfo {volume} {108}},\ \bibinfo
  {pages} {057002} (\bibinfo {year} {2012})}\BibitemShut {NoStop}%
\bibitem [{\citenamefont {Zhang}\ \emph
  {et~al.}(2014{\natexlab{a}})\citenamefont {Zhang}, \citenamefont {Souza},
  \citenamefont {Brandao},\ and\ \citenamefont {Suter}}]{suterprotectedgate}%
  \BibitemOpen
  \bibfield  {author} {\bibinfo {author} {\bibfnamefont {J.}~\bibnamefont
  {Zhang}}, \bibinfo {author} {\bibfnamefont {A.~M.}\ \bibnamefont {Souza}},
  \bibinfo {author} {\bibfnamefont {F.~D.}\ \bibnamefont {Brandao}}, \ and\
  \bibinfo {author} {\bibfnamefont {D.}~\bibnamefont {Suter}},\ }\href@noop {}
  {\bibfield  {journal} {\bibinfo  {journal} {Phys. Rev. Lett.}\ }\textbf
  {\bibinfo {volume} {112}},\ \bibinfo {pages} {050502} (\bibinfo {year}
  {2014}{\natexlab{a}})}\BibitemShut {NoStop}%
\bibitem [{\citenamefont {Howard}\ \emph {et~al.}(2006)\citenamefont {Howard},
  \citenamefont {Twamley}, \citenamefont {Wittmann}, \citenamefont {Gaebel},
  \citenamefont {Jelezko},\ and\ \citenamefont
  {Wrachtrup}}]{howard2006quantum}%
  \BibitemOpen
  \bibfield  {author} {\bibinfo {author} {\bibfnamefont {M.}~\bibnamefont
  {Howard}}, \bibinfo {author} {\bibfnamefont {J.}~\bibnamefont {Twamley}},
  \bibinfo {author} {\bibfnamefont {C.}~\bibnamefont {Wittmann}}, \bibinfo
  {author} {\bibfnamefont {T.}~\bibnamefont {Gaebel}}, \bibinfo {author}
  {\bibfnamefont {F.}~\bibnamefont {Jelezko}}, \ and\ \bibinfo {author}
  {\bibfnamefont {J.}~\bibnamefont {Wrachtrup}},\ }\href@noop {} {\bibfield
  {journal} {\bibinfo  {journal} {New J. Phys.}\ }\textbf {\bibinfo {volume}
  {8}},\ \bibinfo {pages} {33} (\bibinfo {year} {2006})}\BibitemShut {NoStop}%
\bibitem [{\citenamefont {Shukla}\ and\ \citenamefont {Mahesh}(2014)}]{sspt}%
  \BibitemOpen
  \bibfield  {author} {\bibinfo {author} {\bibfnamefont {A.}~\bibnamefont
  {Shukla}}\ and\ \bibinfo {author} {\bibfnamefont {T.}~\bibnamefont
  {Mahesh}},\ }\href@noop {} {\bibfield  {journal} {\bibinfo  {journal} {Phys.
  Rev. A}\ }\textbf {\bibinfo {volume} {90}},\ \bibinfo {pages} {052301}
  (\bibinfo {year} {2014})}\BibitemShut {NoStop}%
\bibitem [{\citenamefont {Pogorelov}\ \emph {et~al.}(2017)\citenamefont
  {Pogorelov}, \citenamefont {Struchalin}, \citenamefont {Straupe},
  \citenamefont {Radchenko}, \citenamefont {Kravtsov},\ and\ \citenamefont
  {Kulik}}]{pogorelov2017experimental}%
  \BibitemOpen
  \bibfield  {author} {\bibinfo {author} {\bibfnamefont {I.}~\bibnamefont
  {Pogorelov}}, \bibinfo {author} {\bibfnamefont {G.}~\bibnamefont
  {Struchalin}}, \bibinfo {author} {\bibfnamefont {S.}~\bibnamefont {Straupe}},
  \bibinfo {author} {\bibfnamefont {I.}~\bibnamefont {Radchenko}}, \bibinfo
  {author} {\bibfnamefont {K.}~\bibnamefont {Kravtsov}}, \ and\ \bibinfo
  {author} {\bibfnamefont {S.}~\bibnamefont {Kulik}},\ }\href@noop {}
  {\bibfield  {journal} {\bibinfo  {journal} {Phys. Rev. A}\ }\textbf {\bibinfo
  {volume} {95}},\ \bibinfo {pages} {012302} (\bibinfo {year}
  {2017})}\BibitemShut {NoStop}%
\bibitem [{\citenamefont {Branderhorst}\ \emph {et~al.}(2009)\citenamefont
  {Branderhorst}, \citenamefont {Nunn}, \citenamefont {Walmsley},\ and\
  \citenamefont {Kosut}}]{simplifiedQPT1}%
  \BibitemOpen
  \bibfield  {author} {\bibinfo {author} {\bibfnamefont {M.}~\bibnamefont
  {Branderhorst}}, \bibinfo {author} {\bibfnamefont {J.}~\bibnamefont {Nunn}},
  \bibinfo {author} {\bibfnamefont {I.}~\bibnamefont {Walmsley}}, \ and\
  \bibinfo {author} {\bibfnamefont {R.}~\bibnamefont {Kosut}},\ }\href@noop {}
  {\bibfield  {journal} {\bibinfo  {journal} {New J. Phys.}\ }\textbf {\bibinfo
  {volume} {11}},\ \bibinfo {pages} {115010} (\bibinfo {year}
  {2009})}\BibitemShut {NoStop}%
\bibitem [{\citenamefont {Wu}\ \emph {et~al.}(2013)\citenamefont {Wu},
  \citenamefont {Li}, \citenamefont {Zheng}, \citenamefont {Peng},\ and\
  \citenamefont {Feng}}]{simplifiedQPT2}%
  \BibitemOpen
  \bibfield  {author} {\bibinfo {author} {\bibfnamefont {Z.}~\bibnamefont
  {Wu}}, \bibinfo {author} {\bibfnamefont {S.}~\bibnamefont {Li}}, \bibinfo
  {author} {\bibfnamefont {W.}~\bibnamefont {Zheng}}, \bibinfo {author}
  {\bibfnamefont {X.}~\bibnamefont {Peng}}, \ and\ \bibinfo {author}
  {\bibfnamefont {M.}~\bibnamefont {Feng}},\ }\href@noop {} {\bibfield
  {journal} {\bibinfo  {journal} {J. Chem. Phys.}\ }\textbf {\bibinfo {volume}
  {138}},\ \bibinfo {pages} {024318} (\bibinfo {year} {2013})}\BibitemShut
  {NoStop}%
\bibitem [{\citenamefont {Dobrovitski}\ \emph {et~al.}(2010)\citenamefont
  {Dobrovitski}, \citenamefont {De~Lange}, \citenamefont {Riste},\ and\
  \citenamefont {Hanson}}]{dobrovitski2010bootstrap}%
  \BibitemOpen
  \bibfield  {author} {\bibinfo {author} {\bibfnamefont {V.}~\bibnamefont
  {Dobrovitski}}, \bibinfo {author} {\bibfnamefont {G.}~\bibnamefont
  {De~Lange}}, \bibinfo {author} {\bibfnamefont {D.}~\bibnamefont {Riste}}, \
  and\ \bibinfo {author} {\bibfnamefont {R.}~\bibnamefont {Hanson}},\
  }\href@noop {} {\bibfield  {journal} {\bibinfo  {journal} {Phys. Rev. Lett.}\
  }\textbf {\bibinfo {volume} {105}},\ \bibinfo {pages} {077601} (\bibinfo
  {year} {2010})}\BibitemShut {NoStop}%
\bibitem [{\citenamefont {Gaikwad}\ \emph {et~al.}(2018)\citenamefont
  {Gaikwad}, \citenamefont {Rehal}, \citenamefont {Singh}, \citenamefont
  {Dorai} \emph {et~al.}}]{gaikwad}%
  \BibitemOpen
  \bibfield  {author} {\bibinfo {author} {\bibfnamefont {A.}~\bibnamefont
  {Gaikwad}}, \bibinfo {author} {\bibfnamefont {D.}~\bibnamefont {Rehal}},
  \bibinfo {author} {\bibfnamefont {A.}~\bibnamefont {Singh}}, \bibinfo
  {author} {\bibfnamefont {K.}~\bibnamefont {Dorai}},  \emph {et~al.},\
  }\href@noop {} {\bibfield  {journal} {\bibinfo  {journal} {Phys. Rev. A}\
  }\textbf {\bibinfo {volume} {97}},\ \bibinfo {pages} {022311} (\bibinfo
  {year} {2018})}\BibitemShut {NoStop}%
\bibitem [{\citenamefont {Mohseni}\ and\ \citenamefont
  {Lidar}(2006)}]{mohseni2006direct}%
  \BibitemOpen
  \bibfield  {author} {\bibinfo {author} {\bibfnamefont {M.}~\bibnamefont
  {Mohseni}}\ and\ \bibinfo {author} {\bibfnamefont {D.}~\bibnamefont
  {Lidar}},\ }\href@noop {} {\bibfield  {journal} {\bibinfo  {journal} {Phys.
  Rev. Lett}\ }\textbf {\bibinfo {volume} {97}},\ \bibinfo {pages} {170501}
  (\bibinfo {year} {2006})}\BibitemShut {NoStop}%
\bibitem [{IBM(2016{\natexlab{a}})}]{IBMQXE}%
  \BibitemOpen
  \href@noop {} {\enquote {\bibinfo {title} {{IBM Quantum Computing
  Platform}},}\ }\bibinfo {howpublished}
  {\url{http://research.ibm.com/ibm-q/qx/}} (\bibinfo {year}
  {2016}{\natexlab{a}}),\ \bibinfo {note} {online accessed
  04-May-2016}\BibitemShut {NoStop}%
\bibitem [{IBM(2016{\natexlab{b}})}]{IBMQX}%
  \BibitemOpen
  \href@noop {} {\enquote {\bibinfo {title} {Ibmqx2 quantum processor},}\
  }\bibinfo {howpublished}
  {\url{https://quantumexperience.ng.bluemix.net/qx/editor}} (\bibinfo {year}
  {2016}{\natexlab{b}})\BibitemShut {NoStop}%
\bibitem [{\citenamefont {Devitt}(2016)}]{devitt2016performing}%
  \BibitemOpen
  \bibfield  {author} {\bibinfo {author} {\bibfnamefont {S.~J.}\ \bibnamefont
  {Devitt}},\ }\href@noop {} {\bibfield  {journal} {\bibinfo  {journal} {Phys.
  Rev. A}\ }\textbf {\bibinfo {volume} {94}},\ \bibinfo {pages} {032329}
  (\bibinfo {year} {2016})}\BibitemShut {NoStop}%
\bibitem [{\citenamefont {Wang}\ \emph {et~al.}(2018)\citenamefont {Wang},
  \citenamefont {Li}, \citenamefont {Yin},\ and\ \citenamefont
  {Zeng}}]{wang201816}%
  \BibitemOpen
  \bibfield  {author} {\bibinfo {author} {\bibfnamefont {Y.}~\bibnamefont
  {Wang}}, \bibinfo {author} {\bibfnamefont {Y.}~\bibnamefont {Li}}, \bibinfo
  {author} {\bibfnamefont {Z.-q.}\ \bibnamefont {Yin}}, \ and\ \bibinfo
  {author} {\bibfnamefont {B.}~\bibnamefont {Zeng}},\ }\href@noop {} {\bibfield
   {journal} {\bibinfo  {journal} {arXiv preprint arXiv:1801.03782}\ }
  (\bibinfo {year} {2018})}\BibitemShut {NoStop}%
\bibitem [{\citenamefont {Kandala}\ \emph {et~al.}(2017)\citenamefont
  {Kandala}, \citenamefont {Mezzacapo}, \citenamefont {Temme}, \citenamefont
  {Takita}, \citenamefont {Brink}, \citenamefont {Chow},\ and\ \citenamefont
  {Gambetta}}]{kandala123}%
  \BibitemOpen
  \bibfield  {author} {\bibinfo {author} {\bibfnamefont {A.}~\bibnamefont
  {Kandala}}, \bibinfo {author} {\bibfnamefont {A.}~\bibnamefont {Mezzacapo}},
  \bibinfo {author} {\bibfnamefont {K.}~\bibnamefont {Temme}}, \bibinfo
  {author} {\bibfnamefont {M.}~\bibnamefont {Takita}}, \bibinfo {author}
  {\bibfnamefont {M.}~\bibnamefont {Brink}}, \bibinfo {author} {\bibfnamefont
  {J.~M.}\ \bibnamefont {Chow}}, \ and\ \bibinfo {author} {\bibfnamefont
  {J.~M.}\ \bibnamefont {Gambetta}},\ }\href@noop {} {\bibfield  {journal}
  {\bibinfo  {journal} {Nature}\ }\textbf {\bibinfo {volume} {549}},\ \bibinfo
  {pages} {242} (\bibinfo {year} {2017})}\BibitemShut {NoStop}%
\bibitem [{\citenamefont {Dall'Arno}\ \emph {et~al.}(2018)\citenamefont
  {Dall'Arno}, \citenamefont {Buscemi},\ and\ \citenamefont
  {Vedral}}]{dall2018device}%
  \BibitemOpen
  \bibfield  {author} {\bibinfo {author} {\bibfnamefont {M.}~\bibnamefont
  {Dall'Arno}}, \bibinfo {author} {\bibfnamefont {F.}~\bibnamefont {Buscemi}},
  \ and\ \bibinfo {author} {\bibfnamefont {V.}~\bibnamefont {Vedral}},\
  }\href@noop {} {\bibfield  {journal} {\bibinfo  {journal} {arXiv preprint
  arXiv:1805.01159}\ } (\bibinfo {year} {2018})}\BibitemShut {NoStop}%
\bibitem [{\citenamefont {Sisodia}\ \emph
  {et~al.}(2017{\natexlab{a}})\citenamefont {Sisodia}, \citenamefont {Shukla},
  \citenamefont {Thapliyal},\ and\ \citenamefont {Pathak}}]{sisodia2017design}%
  \BibitemOpen
  \bibfield  {author} {\bibinfo {author} {\bibfnamefont {M.}~\bibnamefont
  {Sisodia}}, \bibinfo {author} {\bibfnamefont {A.}~\bibnamefont {Shukla}},
  \bibinfo {author} {\bibfnamefont {K.}~\bibnamefont {Thapliyal}}, \ and\
  \bibinfo {author} {\bibfnamefont {A.}~\bibnamefont {Pathak}},\ }\href@noop {}
  {\bibfield  {journal} {\bibinfo  {journal} {Quantum Inf. Process.}\ }\textbf
  {\bibinfo {volume} {16}},\ \bibinfo {pages} {292} (\bibinfo {year}
  {2017}{\natexlab{a}})}\BibitemShut {NoStop}%
\bibitem [{\citenamefont {Sisodia}\ \emph
  {et~al.}(2017{\natexlab{b}})\citenamefont {Sisodia}, \citenamefont {Shukla},\
  and\ \citenamefont {Pathak}}]{abc}%
  \BibitemOpen
  \bibfield  {author} {\bibinfo {author} {\bibfnamefont {M.}~\bibnamefont
  {Sisodia}}, \bibinfo {author} {\bibfnamefont {A.}~\bibnamefont {Shukla}}, \
  and\ \bibinfo {author} {\bibfnamefont {A.}~\bibnamefont {Pathak}},\
  }\href@noop {} {\bibfield  {journal} {\bibinfo  {journal} {Phys. Lett. A}\
  }\textbf {\bibinfo {volume} {381}},\ \bibinfo {pages} {3860} (\bibinfo {year}
  {2017}{\natexlab{b}})}\BibitemShut {NoStop}%
\bibitem [{\citenamefont {Satyajit}\ \emph {et~al.}(2017)\citenamefont
  {Satyajit}, \citenamefont {Srinivasan}, \citenamefont {Behera},\ and\
  \citenamefont {Panigrahi}}]{satyajit2017discrimination}%
  \BibitemOpen
  \bibfield  {author} {\bibinfo {author} {\bibfnamefont {S.}~\bibnamefont
  {Satyajit}}, \bibinfo {author} {\bibfnamefont {K.}~\bibnamefont
  {Srinivasan}}, \bibinfo {author} {\bibfnamefont {B.~K.}\ \bibnamefont
  {Behera}}, \ and\ \bibinfo {author} {\bibfnamefont {P.~K.}\ \bibnamefont
  {Panigrahi}},\ }\href@noop {} {\bibfield  {journal} {\bibinfo  {journal}
  {arXiv preprint arXiv:1712.05485}\ } (\bibinfo {year} {2017})}\BibitemShut
  {NoStop}%
\bibitem [{bac(2017{\natexlab{a}})}]{backendQx2}%
  \BibitemOpen
  \href@noop {} {\enquote {\bibinfo {title} {{IBMQX2 Architecture}},}\
  }\bibinfo {howpublished}
  {"https://github.com/{IBM}/qiskit-qx-info/blob/master/backends/{IBM}qx2/README.md"}
  (\bibinfo {year} {2017}{\natexlab{a}})\BibitemShut {NoStop}%
\bibitem [{\citenamefont {Nielsen}\ and\ \citenamefont
  {Chuang}(2010)}]{chuangbook}%
  \BibitemOpen
  \bibfield  {author} {\bibinfo {author} {\bibfnamefont {M.~A.}\ \bibnamefont
  {Nielsen}}\ and\ \bibinfo {author} {\bibfnamefont {I.~L.}\ \bibnamefont
  {Chuang}},\ }\href@noop {} {\emph {\bibinfo {title} {Quantum computation and
  quantum information}}}\ (\bibinfo  {publisher} {Cambridge university press},\
  \bibinfo {year} {2010})\BibitemShut {NoStop}%
\bibitem [{bac(2017{\natexlab{b}})}]{backendQx4}%
  \BibitemOpen
  \href@noop {} {\enquote {\bibinfo {title} {{IBMQX4 Architecture}},}\
  }\bibinfo {howpublished}
  {"https://github.com/{IBM}/qiskit-qx-info/blob/master/backends/{IBM}qx4/README.md"}
  (\bibinfo {year} {2017}{\natexlab{b}})\BibitemShut {NoStop}%
\bibitem [{\citenamefont {Behera}\ \emph {et~al.}(2017)\citenamefont {Behera},
  \citenamefont {Banerjee},\ and\ \citenamefont
  {Panigrahi}}]{behera2017experimental}%
  \BibitemOpen
  \bibfield  {author} {\bibinfo {author} {\bibfnamefont {B.~K.}\ \bibnamefont
  {Behera}}, \bibinfo {author} {\bibfnamefont {A.}~\bibnamefont {Banerjee}}, \
  and\ \bibinfo {author} {\bibfnamefont {P.~K.}\ \bibnamefont {Panigrahi}},\
  }\href@noop {} {\bibfield  {journal} {\bibinfo  {journal} {Quantum Inf.
  Process.}\ }\textbf {\bibinfo {volume} {16}},\ \bibinfo {pages} {312}
  (\bibinfo {year} {2017})}\BibitemShut {NoStop}%
\bibitem [{\citenamefont {Behera}\ \emph {et~al.}(2018)\citenamefont {Behera},
  \citenamefont {Reza}, \citenamefont {Gupta},\ and\ \citenamefont
  {Panigrahi}}]{behera2018designing}%
  \BibitemOpen
  \bibfield  {author} {\bibinfo {author} {\bibfnamefont {B.~K.}\ \bibnamefont
  {Behera}}, \bibinfo {author} {\bibfnamefont {T.}~\bibnamefont {Reza}},
  \bibinfo {author} {\bibfnamefont {A.}~\bibnamefont {Gupta}}, \ and\ \bibinfo
  {author} {\bibfnamefont {P.~K.}\ \bibnamefont {Panigrahi}},\ }\href@noop {}
  {\bibfield  {journal} {\bibinfo  {journal} {arXiv preprint arXiv:1803.06530}\
  } (\bibinfo {year} {2018})}\BibitemShut {NoStop}%
\bibitem [{\citenamefont {Cory}\ \emph {et~al.}(2000)\citenamefont {Cory},
  \citenamefont {Laflamme}, \citenamefont {Knill}, \citenamefont {Viola},
  \citenamefont {Havel}, \citenamefont {Boulant}, \citenamefont {Boutis},
  \citenamefont {Fortunato}, \citenamefont {Lloyd}, \citenamefont {Martinez}
  \emph {et~al.}}]{cory2000nmr}%
  \BibitemOpen
  \bibfield  {author} {\bibinfo {author} {\bibfnamefont {D.~G.}\ \bibnamefont
  {Cory}}, \bibinfo {author} {\bibfnamefont {R.}~\bibnamefont {Laflamme}},
  \bibinfo {author} {\bibfnamefont {E.}~\bibnamefont {Knill}}, \bibinfo
  {author} {\bibfnamefont {L.}~\bibnamefont {Viola}}, \bibinfo {author}
  {\bibfnamefont {T.}~\bibnamefont {Havel}}, \bibinfo {author} {\bibfnamefont
  {N.}~\bibnamefont {Boulant}}, \bibinfo {author} {\bibfnamefont
  {G.}~\bibnamefont {Boutis}}, \bibinfo {author} {\bibfnamefont
  {E.}~\bibnamefont {Fortunato}}, \bibinfo {author} {\bibfnamefont
  {S.}~\bibnamefont {Lloyd}}, \bibinfo {author} {\bibfnamefont
  {R.}~\bibnamefont {Martinez}},  \emph {et~al.},\ }\href@noop {} {\bibfield
  {journal} {\bibinfo  {journal} {Fortschritte der Physik}\ }\textbf {\bibinfo
  {volume} {48}},\ \bibinfo {pages} {875} (\bibinfo {year} {2000})}\BibitemShut
  {NoStop}%
\bibitem [{\citenamefont {Cory}\ \emph {et~al.}(1997)\citenamefont {Cory},
  \citenamefont {Fahmy},\ and\ \citenamefont {Havel}}]{cory2000nmr1}%
  \BibitemOpen
  \bibfield  {author} {\bibinfo {author} {\bibfnamefont {D.~G.}\ \bibnamefont
  {Cory}}, \bibinfo {author} {\bibfnamefont {A.~F.}\ \bibnamefont {Fahmy}}, \
  and\ \bibinfo {author} {\bibfnamefont {T.~F.}\ \bibnamefont {Havel}},\
  }\href@noop {} {\bibfield  {journal} {\bibinfo  {journal} {Proc. Natl. Acad.
  Sci. USA}\ }\textbf {\bibinfo {volume} {94}},\ \bibinfo {pages} {1634}
  (\bibinfo {year} {1997})}\BibitemShut {NoStop}%
\bibitem [{\citenamefont {Zhang}\ \emph
  {et~al.}(2014{\natexlab{b}})\citenamefont {Zhang}, \citenamefont {Souza},
  \citenamefont {Brandao},\ and\ \citenamefont {Suter}}]{zhang2014protected}%
  \BibitemOpen
  \bibfield  {author} {\bibinfo {author} {\bibfnamefont {J.}~\bibnamefont
  {Zhang}}, \bibinfo {author} {\bibfnamefont {A.~M.}\ \bibnamefont {Souza}},
  \bibinfo {author} {\bibfnamefont {F.~D.}\ \bibnamefont {Brandao}}, \ and\
  \bibinfo {author} {\bibfnamefont {D.}~\bibnamefont {Suter}},\ }\href@noop {}
  {\bibfield  {journal} {\bibinfo  {journal} {Phys. Rev. Lett.}\ }\textbf
  {\bibinfo {volume} {112}},\ \bibinfo {pages} {050502} (\bibinfo {year}
  {2014}{\natexlab{b}})}\BibitemShut {NoStop}%
\bibitem [{\citenamefont {Lucero}\ \emph {et~al.}(2008)\citenamefont {Lucero},
  \citenamefont {Hofheinz}, \citenamefont {Ansmann}, \citenamefont {Bialczak},
  \citenamefont {Katz}, \citenamefont {Neeley}, \citenamefont {O’Connell},
  \citenamefont {Wang}, \citenamefont {Cleland},\ and\ \citenamefont
  {Martinis}}]{lucero2008high}%
  \BibitemOpen
  \bibfield  {author} {\bibinfo {author} {\bibfnamefont {E.}~\bibnamefont
  {Lucero}}, \bibinfo {author} {\bibfnamefont {M.}~\bibnamefont {Hofheinz}},
  \bibinfo {author} {\bibfnamefont {M.}~\bibnamefont {Ansmann}}, \bibinfo
  {author} {\bibfnamefont {R.~C.}\ \bibnamefont {Bialczak}}, \bibinfo {author}
  {\bibfnamefont {N.}~\bibnamefont {Katz}}, \bibinfo {author} {\bibfnamefont
  {M.}~\bibnamefont {Neeley}}, \bibinfo {author} {\bibfnamefont
  {A.}~\bibnamefont {O’Connell}}, \bibinfo {author} {\bibfnamefont
  {H.}~\bibnamefont {Wang}}, \bibinfo {author} {\bibfnamefont {A.}~\bibnamefont
  {Cleland}}, \ and\ \bibinfo {author} {\bibfnamefont {J.~M.}\ \bibnamefont
  {Martinis}},\ }\href@noop {} {\bibfield  {journal} {\bibinfo  {journal}
  {Phys. Rev. Lett.}\ }\textbf {\bibinfo {volume} {100}},\ \bibinfo {pages}
  {247001} (\bibinfo {year} {2008})}\BibitemShut {NoStop}%
\bibitem [{\citenamefont {Doherty}\ \emph {et~al.}(2013)\citenamefont
  {Doherty}, \citenamefont {Manson}, \citenamefont {Delaney}, \citenamefont
  {Jelezko}, \citenamefont {Wrachtrup},\ and\ \citenamefont
  {Hollenberg}}]{doherty2013nitrogen}%
  \BibitemOpen
  \bibfield  {author} {\bibinfo {author} {\bibfnamefont {M.~W.}\ \bibnamefont
  {Doherty}}, \bibinfo {author} {\bibfnamefont {N.~B.}\ \bibnamefont {Manson}},
  \bibinfo {author} {\bibfnamefont {P.}~\bibnamefont {Delaney}}, \bibinfo
  {author} {\bibfnamefont {F.}~\bibnamefont {Jelezko}}, \bibinfo {author}
  {\bibfnamefont {J.}~\bibnamefont {Wrachtrup}}, \ and\ \bibinfo {author}
  {\bibfnamefont {L.~C.}\ \bibnamefont {Hollenberg}},\ }\href@noop {}
  {\bibfield  {journal} {\bibinfo  {journal} {Physics Reports}\ }\textbf
  {\bibinfo {volume} {528}},\ \bibinfo {pages} {1} (\bibinfo {year}
  {2013})}\BibitemShut {NoStop}%
\bibitem [{\citenamefont {Chow}\ \emph
  {et~al.}(2009{\natexlab{b}})\citenamefont {Chow}, \citenamefont {Gambetta},
  \citenamefont {Tornberg}, \citenamefont {Koch}, \citenamefont {Bishop},
  \citenamefont {Houck}, \citenamefont {Johnson}, \citenamefont {Frunzio},
  \citenamefont {Girvin},\ and\ \citenamefont
  {Schoelkopf}}]{chow2009randomized}%
  \BibitemOpen
  \bibfield  {author} {\bibinfo {author} {\bibfnamefont {J.}~\bibnamefont
  {Chow}}, \bibinfo {author} {\bibfnamefont {J.~M.}\ \bibnamefont {Gambetta}},
  \bibinfo {author} {\bibfnamefont {L.}~\bibnamefont {Tornberg}}, \bibinfo
  {author} {\bibfnamefont {J.}~\bibnamefont {Koch}}, \bibinfo {author}
  {\bibfnamefont {L.~S.}\ \bibnamefont {Bishop}}, \bibinfo {author}
  {\bibfnamefont {A.~A.}\ \bibnamefont {Houck}}, \bibinfo {author}
  {\bibfnamefont {B.}~\bibnamefont {Johnson}}, \bibinfo {author} {\bibfnamefont
  {L.}~\bibnamefont {Frunzio}}, \bibinfo {author} {\bibfnamefont {S.~M.}\
  \bibnamefont {Girvin}}, \ and\ \bibinfo {author} {\bibfnamefont {R.~J.}\
  \bibnamefont {Schoelkopf}},\ }\href@noop {} {\bibfield  {journal} {\bibinfo
  {journal} {Phys. Rev. Lett.}\ }\textbf {\bibinfo {volume} {102}},\ \bibinfo
  {pages} {090502} (\bibinfo {year} {2009}{\natexlab{b}})}\BibitemShut
  {NoStop}%
\bibitem [{\citenamefont {Mohseni}\ \emph {et~al.}(2008)\citenamefont
  {Mohseni}, \citenamefont {Rezakhani},\ and\ \citenamefont
  {Lidar}}]{mohseni2008quantum}%
  \BibitemOpen
  \bibfield  {author} {\bibinfo {author} {\bibfnamefont {M.}~\bibnamefont
  {Mohseni}}, \bibinfo {author} {\bibfnamefont {A.}~\bibnamefont {Rezakhani}},
  \ and\ \bibinfo {author} {\bibfnamefont {D.}~\bibnamefont {Lidar}},\
  }\href@noop {} {\bibfield  {journal} {\bibinfo  {journal} {Phys. Rev. A}\
  }\textbf {\bibinfo {volume} {77}},\ \bibinfo {pages} {032322} (\bibinfo
  {year} {2008})}\BibitemShut {NoStop}%
\end{thebibliography}%
\end{document}